\documentclass[superscriptaddress,aps,prb,reprint,twocolumn,amsmath,amssymb]{revtex4-2}
\usepackage{graphicx}
\usepackage{hyperref}
\usepackage{amssymb}
\usepackage{slashed}
\usepackage{dcolumn}
\usepackage{amsmath}
\usepackage{bm}
\usepackage{colordvi}
\usepackage{algorithm}
\usepackage{algpseudocode}
\usepackage{multirow}
\usepackage{titlesec}
\usepackage{newtxtext,newtxmath}
\usepackage{dsfont}
\usepackage{xcolor}
\usepackage{mathbbol}
\usepackage{appendix}

\usepackage{mathrsfs}
\makeatletter

\newcommand{\Rmnum}[1]{\expandafter\@slowromancap\romannumeral #1@}
\makeatother

\begin{document}   

\title{A Self-Consistent Computational Framework  for Displacive Ferroelectrics from the Condensed Ground State}   

\author{Fei Yang}
\email{fzy5099@psu.edu}

\affiliation{Department of Materials Science and Engineering and Materials Research Institute, The Pennsylvania State University, University Park, PA 16802, USA}

\author{Long-Qing Chen}
\email{lqc3@psu.edu}

\affiliation{Department of Materials Science and Engineering and Materials Research Institute, The Pennsylvania State University, University Park, PA 16802, USA}

\date{\today}

\begin{abstract}
Quantitative description of finite-temperature properties of displacive ferroelectrics, and in particular the critical behavior, is of fundamental importance to both theory and device design, going beyond the Landau-Ginzburg approach, which requires known knowledge of critical behaviors and temperature-dependent parameter fitting. Here within quantum statistic description of polarization fluctuations, we develop a self-consistent, microscopically based computational
 framework for finite-temperature thermodynamics and phase transitions in displacive ferroelectrics. It enables one to use only the ground-state properties to predict the finite-temperature properties and in particular, the criticality of phase transitions of various displacive ferroelectrics.  Its applications to the classical ferroelectric PbTiO$_3$, quantum paraelectrics SrTiO$_3$ and KTaO$_3$, and recently fabricated  ferroelectric strained SrTiO$_3$, demonstrate remarkable quantitative agreements with the experimentally measured dielectric/ferroelectric properties throughout the entire temperature ranges of the phases, including the critical behaviors of phase transitions. The proposed computational framework offers a tractable  quantitative  basis for bridging microscopic ground-state modeling and macroscopic device-level design in a broad range of ferroelectric systems under diverse thermodynamic and external conditions.
\end{abstract}


\maketitle  

\section*{Introduction}

Existing descriptions for thermodynamics of displacive ferroelectrics have largely relied on the phenomenological Landau-Ginzburg  theory~\cite{lines2001principles,devonshire1954theory,haun1989thermodynamic,haun1987thermodynamic,chen2003nonlinear}. This approach expands the free energy around critical point, and has been successful in describing general behaviors of  ferroelectric properties near phase transitions. However, for a quantitative description of a given ferroelectric system, it requires the knowledge of the known critical behaviors (e.g., critical temperature and the type of transition) as well as polarization  behaviors from experiments to fit model parameters~\cite{lines2001principles,devonshire1954theory,haun1989thermodynamic,haun1987thermodynamic,chen2003nonlinear}, and often has poor description of the ferroelectric properties at low temperatures far away from critical point. There have also been several theoretical attempts in developing a microscopic thermodynamic formalism in displacive ferroelectrics~\cite{rowley2014ferroelectric,conduit2010theory,roussev2003theory,palova2009quantum,silverman1963temperature,rechester1971contribution}, but the predictions have remained largely qualitative. The computational approaches have therefore relied heavily on Monte Carlo simulations constructed from first-principles-based calculations, such as effective Hamiltonians~\cite{tadmor2002polarization,zhong1994phase,zhong1995first,waghmare1997ab,bellaiche2000finite,naumov2004unusual,kornev2006phase,akbarzadeh2012finite}, self-consistent lattice-dynamical calculations~\cite{ghosez1998ab,cohen1992origin,he2020anharmonic,verdi2023quantum}, and empirical or machine-learned interatomic potentials~\cite{qi2016atomistic,gigli2024modeling,wu2022large}. The first-principles calculations have been extremely useful for understanding structural stability and ground state of a ferroelectric crystal, but the finite-temperature  simulations often failed to provide accurate and consistent predictions of properties in the entire temperature range of the phase, particularly the transition temperatures and critical behaviors~\cite{gigli2024modeling,verdi2023quantum}, even when employing state-of-the-art machine-learned interatomic potentials or deep potential molecular dynamics simulations~\cite{PhysRevLett.121.255901,PhysRevB.111.094113}. 

The microscopic quasiparticle formalism of condensed-matter quantum statistical theory~\cite{abrikosov2012methods} allows one to determine the finite-temperature properties and critical behavior of a system solely from the knowledge of its zero-temperature ground state, in contrast to Landau–Ginzburg approach.   Quasiparticle excitations
emerge in an ordered quantum phase associated with a spontaneous breaking of the continuous symmetry upon condensation~\cite{nambu2009nobel}. These excitations determine the macroscopic characteristics and dynamical properties of the system~\cite{abrikosov2012methods}, including the nature of its phase transition and criticality. This framework has been successful in describing ordered quantum systems such as ferromagnets, antiferromagnets, superconductors~\cite{coleman2015introduction,abrikosov2012methods}, and charge-density waves~\cite{lee1974conductivity}, where the ground-state microscopic Hamiltonians (e.g., Heisenberg or BCS-type models) yield quasiparticle excitations (magnons or Bogoliubov quasiparticles) whose thermal populations self-consistently reproduce the finite-temperature phase behavior and the critical phenomena.

A microscopic and self-consistent quasiparticle framework has not been well developed for formulating the finite-temperature {properties} and phase transitions of displacive ferroelectricity, where the order parameter originates from collective lattice displacements (a phonon condensate) rather than from electronic or spin degrees of freedom. In displacive ferroelectrics, upon condensation, a natural identification of the fundamental quasiparticle excitation is the collective amplitude mode of the order parameter, i.e., amplitude fluctuations of the long-range ordered polarization (condensate)~\cite{tang2022excitations,rowley2014ferroelectric,conduit2010theory,roussev2003theory,palova2009quantum,silverman1963temperature,lines2001principles,rechester1971contribution}, as this mode in literature represents the condensed-matter analogue of the Higgs particle in particle physics~\cite{higgs1964broken,englert1964broken} and has been observed in a wide range of systems~\cite{pekker2015amplitude}, including conventional~\cite{matsunaga2013higgs,matsunaga2014light,yang2019gauge,shimano2020higgs} and unconventional~\cite{chu2020phase,yang2020theory} superconductors, superfluids~\cite{pollet2012higgs,endres2012higgs}, certain antiferromagnets~\cite{ruegg2008quantum,jain2017higgs,hong2017higgs}, and charge-density-wave compounds~\cite{yoshikawa2021ultrafast,sugai2006phason,torchinsky2013fluctuating}.  However, in real perovskite ferroelectrics, the polarization and its dynamics are vector (rather than scalar) quantities. Incorporating this vectorial nature of the polarization and
its fluctuations, here we consider the bosonic condensation of the unstable phonons of a displacive ferroelectric crystal, and treat the emerging collective {\sl vectorial} polar mode, i.e., the vector fluctuation of the long-range ordered polarization, as the essential microscopic quasiparticles. By deriving the excitation of this mode with bosonic nature, we develop a self-consistent renormalization from the ground-state Hamiltonian to finite-temperature free energy of the long-range ordered (global) polarization. This offers a self-consistent, microscopically based phase-transition framework of the displacive ferroelectricity, and enables one to use only the ground-state parameters to predict the dielectric/ferroelectric properties at finite temperatures in the entire range of the phase including the criticality of the phase. Without adjusting parameters, its applications to three types of representative materials systems: the classical ferroelectric PbTiO$_3$, quantum paraelectrics SrTiO$_3$ and KTaO$_3$, and recently fabricated ferroelectric strained SrTiO$_3$, demonstrate remarkable quantitative agreements with  experimental measurements,  showing a broad applicability. 

\section*{Model}
\label{secmodel}
The displacive ferroelectricity originates from an unstable transverse optical soft phonon mode~\cite{cochran1981soft,cochran1961crystal,yelon1971neutron,cowley1996phase,cowley1965theory,cochran1969dynamical,cochran1960crystal}, whereas the longitudinal optical mode is strongly stiffened by the long-range dipole–dipole interaction and hence remains inactive. We thus start from the effective lattice-dynamical Hamiltonian describing the transverse soft phonon mode,
\begin{eqnarray}
H&=&\frac{1}{2}\int{d{\bf x}}{d{\bf x'}}\omega^2_{\rm sp}({\bf x},{\bf x}')\phi^*_{\rm sp}({\bf x})\phi_{\rm sp}({\bf x}')\nonumber\\
&&\mbox{}+\int{d{\bf x}}\Big[\frac{u_b}{4}|\phi_{\rm sp}({\bf x})|^4+\frac{u_d}{6}|\phi_{\rm sp}({\bf x})|^6\Big]+H_{\rm C}[\phi_{\rm sp}],~~
\end{eqnarray}
where $\phi_{\rm sp}({\bf x})$ denotes the field operator associated with the soft-phonon mode. The first harmonic term determines the phonon dispersion, with the phonon kernel, 
\begin{equation}
  \omega^2_{\rm sp}({\bf x},{\bf x}')=\Delta^2_{\rm sp}\delta({\bf x-x'})+Q({\bf x-x'}),
\end{equation}
and the second term describes the local self-interaction with coupling strengths $u_b$ and $u_d$.  Here, $\Delta_{\rm sp}$ denotes the excitation gap of the transverse soft phonon mode; $Q({\bf x}-{\bf x'})$ encodes the nonlocal elastic restoring interaction responsible for the phonon dispersion, with Fourier transform $Q({\bf q}) = v^2 q^2$, where ${\bf q}$ is the phonon wavevector in reciprocal space and $v$ is the mode velocity. The coupling effect $H_{\rm C}$ to other phonon branches (e.g., acoustic phonons via  electrostrictive~\cite{palova2009quantum,khmelnitskii1973phase,rowley2014ferroelectric,fujishita2016quantum} or flexoelectric~\cite{PhysRevLett.131.046801} coupling) can be introduced in the form of perturbations or self-energy corrections~\cite{abrikosov2012methods,coleman2015introduction}, as well established in many-body physics.  

{The lattice distortion has been described within the phonon-condensation picture originally introduced by Lee, Rice, and Anderson~\cite{lee1974conductivity}, which also places displacive ferroelectrics in the universality class of bosonic condensation of unstable optical phonons. Specifically, once $\omega^2_{\rm sp}({\bf q}=0)=\Delta^2_{\rm sp}$ becomes negative, the transverse optical  soft phonon mode acquires an imaginary frequency at the Brillouin-zone center ($\Gamma$ point). This signals the onset of a bosonic condensation of the soft-phonon field at $q=0$,  i.e., all imaginary-frequency soft-phonon states collapse into a single macroscopic state at $\Gamma$, exactly analogous to a bosonic condensate. In the mean field theory, such a collective condensation leads to a nonzero expectation value of macroscopic displacement field~\cite{lee1974conductivity},  
  \begin{equation} {\hat \xi}=\langle|{\phi_{\rm sp}(q=0)}|\rangle, 
  \end{equation} 
which describes a global lattice distortion~\cite{lee1974conductivity} and results in a polarization field ${\bf \hat P}=({z^*{\bf e}_{\bf u}}/{\Omega_{\rm cell}}){\hat \xi}$~\cite{cochran1981soft,cochran1961crystal,cowley1996phase,cowley1965theory,cochran1969dynamical,cochran1960crystal,yelon1971neutron}, with ${\bf e}_\mathbf{u}$ the ionic vibration vector, $z^*$ the effective charge, and $\Omega_{\rm cell}$ the unit cell volume. {Consequently, the imaginary-frequency phonon condensation corresponds to the emergence of a macroscopic coherent displacement field, analogous to the order parameter in Bose condensation, but physically realized as a lattice instability rather than particle number condensation. This also justifies focusing on the long-wavelength limit, where the condensation occurs at $q=0$, allowing the theory to be consistently expressed in terms of the macroscopic polarization field.}  Consequently, directly substituting these relations provides a direct connection between the soft-phonon Hamiltonian and the ground-state mean-field Hamiltonian for the long-range polarization field
~\cite{rowley2014ferroelectric,conduit2010theory,roussev2003theory,palova2009quantum}:}
\begin{equation}
\mathcal{H}=\int{d{\bf x}}\Big[\frac{g}{2}({\bf \nabla}{\hat P})^2+\frac{a}{2}{\hat P}^2+\frac{b}{4}{\hat P}^4+\frac{\lambda}{6}{\hat P}^6\Big]+H_{\rm C}[{\hat P}],
\end{equation}  
{where $a\propto\Delta_{\rm sp}^2$, $b\propto{u_b}$, $\lambda\propto{u_d}$ and $g\propto{v^2}$ are model parameters}. It is worth noting that, unlike in superconductors or ferromagnets, whose microscopic ground-state theories (such as the BCS or Heisenberg models)  reduce to a polynomial form only after coarse-graining near the critical temperature $T_c$, the ground-state mean-field  Hamiltonian of a displacive ferroelectric is intrinsically polynomial. It is derived directly from the lattice dynamics of a crystal exhibiting a structural instability, representing a first-principles–inspired model for the unstable phonon mode. This makes it fundamentally distinct from the conventional Ginzburg–Landau free energy, which is constructed phenomenologically as a series expansion near $T_c$ under the assumption of small order parameters.

The polarization field ${\bf \hat P}={\bf P}+\delta{\bf P}$, consisting of the long-range ordered (globally condensed) polarization  ${\bf P}$ and long-wavelength {\sl vectorial}  polarization fluctuation $\delta{\bf P}$ (low-lying excitations),  with a vanishing thermally averaged $\langle\delta{\bf P}\rangle$ but a nonzero $\langle{\delta{P^2}}\rangle$, which can be derived within the quantum statistic mechanism via various approaches (see Appendix). For example, via the derived  equation of motion 
\begin{equation}
\big[m_p\partial_t^2+\gamma\partial_t-g\nabla^2+m_p\Delta^2\big(a,b,P^2,\langle\delta{P}^2\rangle\big)\big]\delta{\bf P}={\bf E}_{\rm th}(t,{\bf R}), 
\end{equation}
where the introduced thermal field ${\bf E}_{\rm th}(t,{\bf R})$ obeys the fluctuation-dissipation theorem~\cite{landau1981statistical}: $\langle{\bf E}_{\rm th}\rangle=0$ and 
\begin{equation}
\langle{\bf E}_{\rm th}(\omega,{\bf q}){\bf E}_{\rm th}(\omega',{\bf q}')\rangle\!=\!\frac{\gamma{\hslash}\omega(2\pi)^4\delta({\bf q}-{\bf q'})\delta(\omega-\omega')}{\tanh[\hslash\omega/(2k_BT)]},  
\end{equation}
one obtains 
\begin{equation}\label{PF}
  \langle{\delta{P}^2}\rangle\!=\!\!\int\!\frac{{\hslash}^2 d{\bf q}}{(2\pi)^3}\frac{2n_B\big[\hslash\omega_{vm}\big(q,a,b,P^2,\langle\delta{P^2}\rangle\big)\big]\!+\!1}{2m_p\hslash\omega_{vm}\big(q,a,b,P^2,\langle\delta{P^2}\rangle\big)},  
\end{equation}
which is related to the microscopic bosonic excitation of the collective {\sl vectorial} polar mode that emerges after the condensation of the imaginary-frequency phonons. {Here, $m_p$ is the polarization inertia, i.e., the inertia coefficient associated with the dynamics of the polarization field, which is given by $m_p=\frac{\Omega_{\rm cell}}{\sum_iQ_i^2/M_i}$~\cite{tang2022excitations,sivasubramanian2004physical,cheng2023terahertz},  with $M_i$ the ionic masses and $Q_i$ the ionic charges in the unit cell.} This quasiparticle excitation has an energy spectrum {\small $\hslash\omega_{vm}\big(q,a,b,P^2,\langle\delta{P^2}\rangle\big)=\hslash\sqrt{\Delta^2\big(a,b,P^2,\langle\delta{P^2}\rangle\big)+gq^2/m_p}$} with $q$ the wave vector and the excitation gap $\Delta$ is determined by self-consistently solving 
\begin{eqnarray}
&&\!\!\!\!\!\!m_p\Delta^2\!=\!a\!+\!b(2/d_p\!+\!1){P}^2\!+\!\lambda(4/d_p\!+\!1){P}^4\!+\!b(2/d_p\!+\!1)\langle\delta{P}^2\rangle \nonumber\\
&&\!\!\!\!\!\!+3\lambda(4/d_p\!+\!1)\langle\delta{P}^2\rangle^2\!+\!2\lambda(4/d_p\!+\!1)(2/d_p\!+\!1){P}^2\langle\delta{P}^2\rangle,\quad\label{gap}
\end{eqnarray}
 due to its dependence on $\langle{\delta{P^2}}\rangle$ and $P^2$. Here, $n_B(x)$ denotes the Bose distribution; $d_p$ is the dimension of the vector space of $\delta{\bf P}${, i.e., the number of fluctuation components retained in the description of $\delta{\bf P}$. While symmetry-reduced descriptions (e.g., $d_p=1$) are sometimes used, retaining $d_p=3$ ensures full fluctuation space and avoids artificial constraints near criticality. Here we consistently retain all three components ({\sl vectorial} polar mode), and $d_p=3$ throughout this work}.

\begin{figure}[htb]
  {\includegraphics[width=8.0cm]{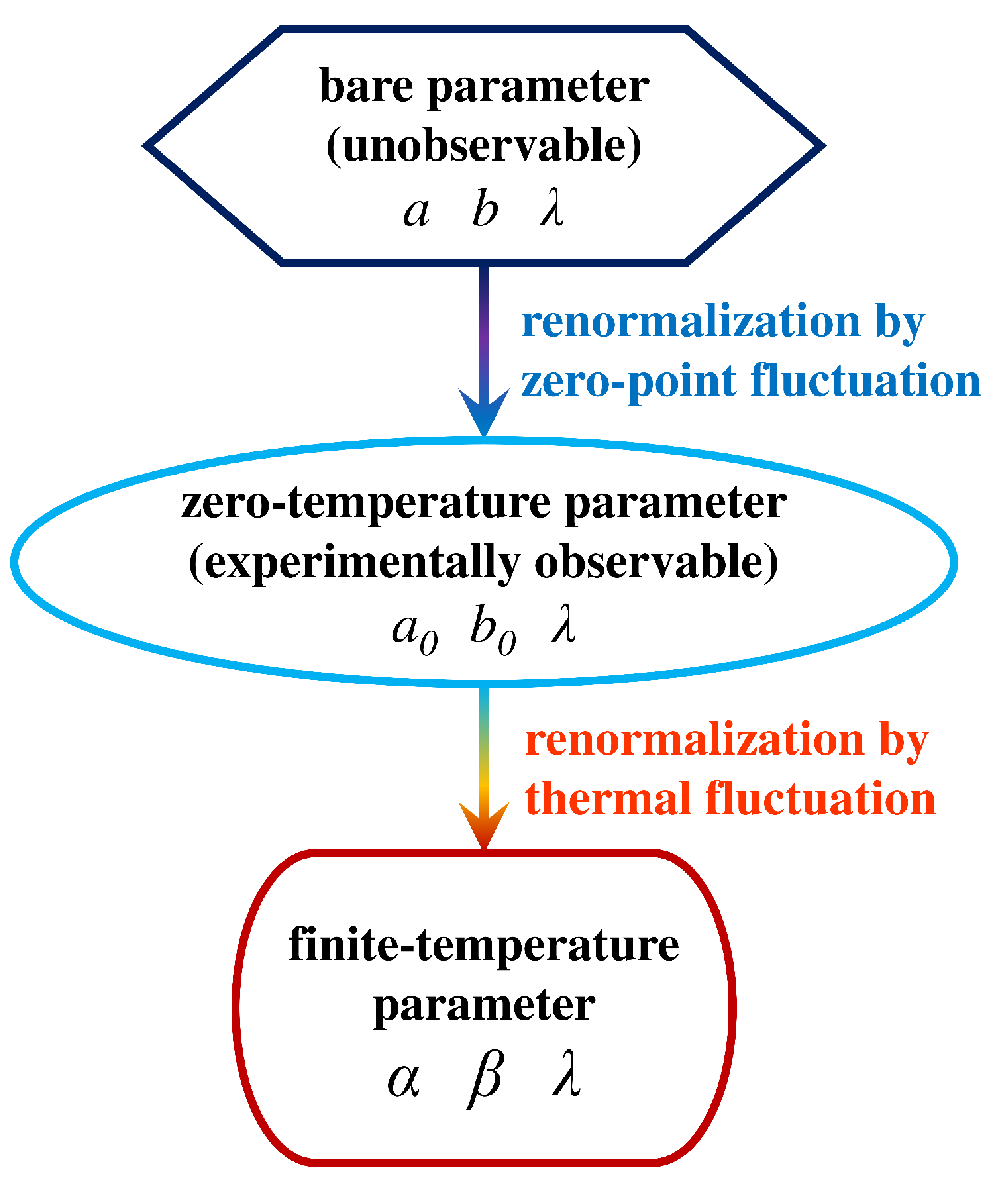}}
  \caption{{{Schematic of the renormalization in thermodynamic theory.  }} }    
\label{Renormalization}
\end{figure}

Due to the bosonic nature of the polar mode, $\langle\delta{P}^2\rangle$ in Eq.~(\ref{PF}) consists of the zero-point fluctuations $\langle\delta{P}^2_{\rm zo}\rangle$ and the thermal fluctuations $\langle\delta{P}^2_{\rm th}(T)\rangle\propto{n_B(\hslash\omega_{\rm vm})}$. The thermal fluctuations are zero at $T=0$ and become finite at $T\ne0$. The zero-point fluctuations as quantum fluctuations are nonzero at $T=0$. In the quantum field theory, the zero-point fluctuations are integrated to the vacuum state~\cite{peskin2018introduction}, and hence, the vacuum state is not truly empty as the original/bare one but instead contains various excitations that pop into and out of the vacuum.  Following this idea, as shown in Fig.~\ref{Renormalization}, we consider that the ground-state parameters $a$ and $b$, which can in principle be obtained from the first-principles calculations,  are experimentally unobservable. Only through the renormalization by zero-point fluctuations $\langle\delta{P}^2_{\rm zo}\rangle$, $a$ and $b$ become the zero-temperature parameters $a_0$ and $b_0$ that can be experimentally measured, yielding the Hamiltonian of the zero-temperature global polarization $P_0$:
\begin{equation}\label{Ham}
\mathcal{\bar H}_0=\int{d{\bf x}}\Big[\frac{a_0}{2}P_0^2+\frac{b_0}{4}P_0^4+\frac{\lambda}{6}P_0^6\Big],
\end{equation}  
where
\begin{eqnarray}
&&\!\!\!\!\!\!a_0\!=\!a\!+\!b(2/d_p\!+\!1)\langle\delta{P}^2_{\rm zo}\rangle\!+\!3\lambda(4/d_p\!+\!1)\big(\langle\delta{P}^2_{\rm zo}\rangle\big)^2\!+\!\delta{a},\quad~\label{a0}\\
&&\!\!\!\!\!\!b_0\!=\!b\!+\!2\lambda(4/d_p\!+\!1)\langle\delta{P}^2_{\rm zo}\rangle, \label{b0}
\end{eqnarray}
 and the zero-point fluctuations:
\begin{equation}\label{zo}
 \langle{\delta{P}^2_{\rm zo}}\rangle=\int\frac{\hslash^2d{\bf q}}{(2\pi)^3}\frac{1}{2m_p\hslash\omega_{vm}(q,a,b,P_0^2,\langle\delta{P_{\rm zo}^2}\rangle)}.  
\end{equation}
The zero-point  polarization $P_0^2$ is determined by the minimum of $\mathcal{\bar H}_0$ self-consistently. Further applying  renormalization by thermal fluctuations $\langle\delta{P}^2_{\rm th}(T)\rangle$ leads to the free energy of global polarization $P(T)$  at finite temperatures, 
\begin{equation}\label{freeenergy}
\mathcal{F}(T)=\int{d{\bf r}}\Big[\frac{\alpha{(T)}}{2}P^2+\frac{\beta(T)}{4}P^4+\frac{\lambda}{6}P^6\Big],
\end{equation}
where the free-energy parameters $\alpha(T)$ and $\beta(T)$:
\begin{eqnarray}
  &&\!\!\!\!\!\!\alpha\!=\!a_0\!+\!b_0(2/d_p\!+\!1)\langle\delta{P}^2_{\rm th}\rangle\!+\!3\lambda(4/d_p\!+\!1)(\langle\delta{P}^2_{\rm th}\rangle)^2\!+\!\delta{\alpha},~~~~~\label{alpha}\\
  &&\!\!\!\!\!\!\beta\!=\!b_0+2\lambda(4/d_p\!+\!1)\langle\delta{P}^2_{\rm th}\rangle,\label{beta}     
\end{eqnarray}
and the thermal fluctuations
\begin{equation}\label{theq}
\langle{\delta{P}^2_{\rm th}}\rangle=\int\frac{\hslash^2d{\bf q}}{(2\pi)^3}\frac{n_B\big[\hslash\omega_{vm}\big(q,a_0,b_0,P^2,\langle\delta{P_{\rm th}^2}\rangle\big)\big]}{m_p\omega_{vm}\big(q,a_0,b_0,P^2,\langle\delta{P_{\rm th}^2}\rangle\big)}.   
\end{equation}  
The minimum of $\mathcal{F}(T)$ gives rise to $P(T)\ne0$ for $\alpha<0$ representing the ferroelectric phase and $P(T)=0$ for $\alpha>0$ representing the paraelectric phase. Consequently, the original parameters $a$ and $b$ in the ground state are renormalized to $\alpha(T)$ and $\beta(T)$ for finite temperatures. {It is noted that we have introduced a correction/perturbation term, $\delta{a}$ in Eq.~(\ref{a0}) and $\delta\alpha(T)$ in Eq.~(\ref{alpha}). These terms represent subleading corrections that account for the  residual or possible   perturbative couplings of the polarization field to other phonon branches (e.g., acoustic phonons), as well as self-consistent structural
instabilities around the transition, which are not explicitly included in the minimal soft-mode description.
In other words, these terms correspond to higher-order self-energy corrections to the soft mode, and are included here for completeness and rigor, with subleading impact on the main results.}  

The self-consistent renormalization of vectorial fluctuations on the ground state here, i.e., self-consistent formulation of the vectorial  fluctuations and order parameter (global polarization) starting from the ground state   enables one to use only the
ground-state parameters to predict the dielectric/ferroelectric
properties at finite temperatures in the entire range of the phase, including the criticality of the phase. For accuracy, one can directly start with the experimentally measured zero-temperature parameters if they are available and perform the renormalization by thermal fluctuations to obtain the finite-temperature dielectric properties. However, if only the bare parameters are available from the first-principles calculations, the renormalization by zero-point fluctuations is required and critical for the accurate predictions of the dielectric properties and criticality. To demonstrate this viewpoint, in the following, we apply the framework to three types of representative materials systems.

\begin{widetext}
\begin{center}
\begin{figure}[htb]
  {\includegraphics[width=17.6cm]{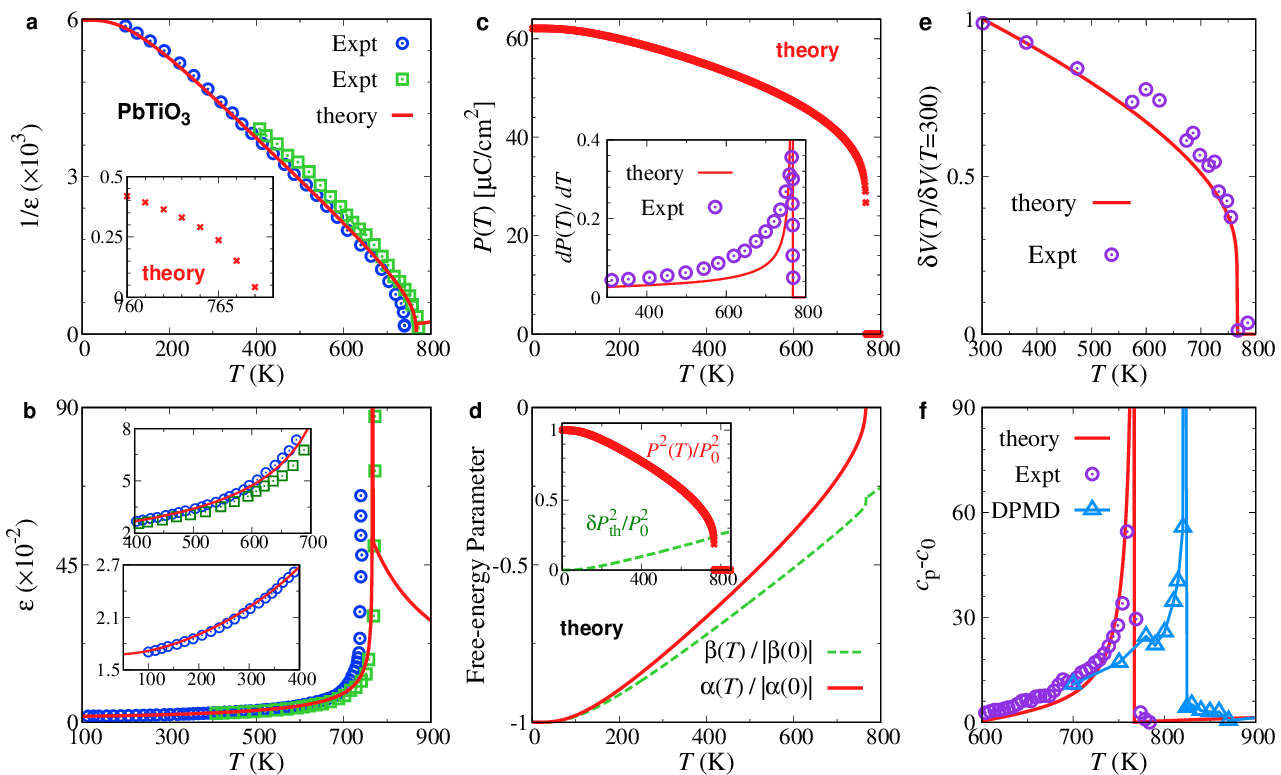}}
  \caption{{{Temperature dependence of the dielectric properties in ferroelectric PbTiO$_3$. {\bf a}, inverse dielectric function; {\bf b}, dielectric function; {\bf c}, spontaneous polarization (order parameter); {\bf d}, free-energy parameters $\alpha(T)$ and $\beta(T)$; {\bf e}, normalized change of unit cell volume (tetragonality); {\bf f}, difference between the calculated specific heat $c_p$ and the Dulong-Petit specific heat $c_0=3nR$ for bulk PbTiO$_3$ with $n=5$~\cite{PhysRevB.111.094113}. Experimental data in {\bf a} and {\bf b} are from Ref.~\cite{remeika1970growth} (squares) and Ref.~\cite{ikegami1971electromechanical} (circles).   
  Insets of {\bf a} and {\bf b} zoom the corresponding temperature ranges. Inset of {\bf c} shows the results of $d{P}/dT$, and the experimental data are from Ref.~\cite{remeika1970growth}. Inset in {\bf d} shows the fluctuation $\langle\delta{P^2_{\rm th}}(T)\rangle$ and order parameter $P^2$. Experimental data in {\bf e} and {\bf f} are from Ref.~\cite{Shirane1951} and Ref.~\cite{Yoshida1960}, respectively. In {\bf f}, for comparison, we also addressed the numerical results from deep potential molecular dynamics (DPMD, with $T_c\approx821~$K) trained on DFT-based data~\cite{PhysRevB.111.094113}.}}}    
\label{fig:PTO}
\end{figure}
\end{center}
\end{widetext}

\section*{Application to classical ferroelectric PbTiO$_3$}

 We first consider the  classical ferroelectric PbTiO$_3$ which has a high transition temperature. The thermal excitation of the collective vectorial polar mode is expected to reduce the ferroelectric polarization with an increase in temperature from zero, thereby leading to the phase transition at certain critical temperature.  As its crystal structure is tetragonal with a large distortion ratio $1.062$~\cite{lines2001principles,samara1971pressure}, we re-write the Hamiltonian as the one limited by crystal symmetry: 
\begin{align}
  &\mathcal{H}\!=\!\!\int{d{\bf x}}\Big[\frac{g}{2}({\nabla}{\bf P})^2\!+\!\frac{a_1}{2}(P_1^2\!+\!P_2^2\!+\!P_3^2)\!+\!\frac{a_{11}}{4}(P_1^4\!+\!P_2^4\!+\!P_3^4)\nonumber\\
  &+\!\frac{a_{111}}{6}(P_1^6\!+\!P_2^6\!+\!P_3^6)\!+\!{a_{12}}(P_1^2P_2^2\!+\!P_2^2P_3^2\!+\!P_1^2P_3^2)\!+\!a_{123}P_1^2P_2^2P_3^2\nonumber\\
 &+a_{112}[P_1^4(P_2^2\!+\!P_3^2)\!+\!P_2^4(P_1^2\!+\!P_3^2)\!+\!P_3^4(P_1^2\!+\!P_2^2)]\Big].\label{PPham}
\end{align}
Similarly, the polarization field ${\bf P}={\bf P}+\delta{\bf P}$, consisting of the long-range ordered polarization ${\bf P}=P{\bf e}_3$ and polarization fluctuation $\delta{\bf P}=({\delta}P_1{\bf e}_1+{\delta}P_2{\bf e}_2+{\delta}P_3{\bf e}_3)/\sqrt{3}$. Following the  derivation procedure of the self-consistent renormalization in Fig.~\ref{Renormalization} (see Appendix), one can derive the free energy of long-range ordered polarization of $P$ (magnitude),
\begin{equation}\label{PPFE}
\mathcal{F}(T)=\int{d{\bf r}}\Big[\frac{\alpha{(T)}}{2}P^2+\frac{\beta(T)}{4}P^4+\frac{\lambda}{6}P^6\Big],
\end{equation}
and use it to calculate the finite-temperature dielectric behavior, the phase evolution of the order parameter (polarization) with increasing temperature, and in particular, the phase-transition criticalities. To start, we directly use the experimentally measured zero-temperature parameters and apply the renormalization by the thermal polarization fluctuations, yielding the free-energy parameters  $\alpha(T)$ and $\beta(T)$. {The specific model parameters $a_{i}$, $a_{ij}$ and $a_{ijk}$ used in our simulation, and their determination using several independent measurements~\cite{haun1987thermodynamic,remeika1970growth,ikegami1971electromechanical} of the spontaneous polarization and dielectric function at low-temperature limit are discussed in Appendix [See  Table~SI for PbTiO$_3$ in Appendix].}

The ferroelectric properties of PbTiO$_3$ predicted from the present theoretical framework are plotted in Fig.~\ref{fig:PTO}. As shown in Fig.~\ref{fig:PTO}a for the inverse dielectric function $1/\varepsilon(T)$ or in Fig.~\ref{fig:PTO}b for  the dielectric function $\varepsilon(T)$, their temperature dependencies closely reproduces the experimental data, showing excellent quantitative agreement across a very wide temperature range (starting from $T=0$ to over $700$~K). The predicted transition temperature $T_c=767~$K (the inset of Fig.~\ref{fig:PTO}a), in excellent agreement with existing experimentally measured value of $763~$K~\cite{lines2001principles,samara1971pressure}. Moreover, according to the predicted temperature dependence of the long-range ordered polarization in Fig.~\ref{fig:PTO}c, the ferroelectric-paraelectric transition in PbTiO$_3$ is first-order, consistent with the widely reported experimental observation~\cite{lines2001principles,samara1971pressure}. Particularly, the obtained $\beta(T=300~\text{K})=0.08175\beta(0)=-0.2943\times10^{9}~$Jm$^5$/C$^4$ from our framework (Fig.~\ref{fig:PTO}d) is almost same as the  measured room-temperature value of $\beta=-0.29\times10^{9}~$Jm$^5$/C$^4$ which has been widely used in the literature~\cite{tang2022excitations,chen2003nonlinear,haun1989thermodynamic,haun1987thermodynamic}.  These remarkable quantitative agreements between theory predictions and various  available experimental data in the ferroelectric PbTiO$_3$ are achieved using only the zero-temperature parameters from existing independent measurements at low-temperature without adjusting parameters or $T$-dependent tuning.

To understand the phase evolution of the polarization with increasing temperature and the criticality of classical ferroelectric phase transition, we plot the free-energy 
 parameters $\alpha(T)$ and $\beta(T)$ in Fig.~\ref{fig:PTO}d. Staring from the zero-temperature parameters $\alpha(0)<0$ and $\beta(0)<0$, the correspondingly renormalized $\alpha(T)$ and $\beta(T)$ increase with temperature. The dielectric properties below $100$~K are insensitive to temperature due to the minimal thermal fluctuation [$n_B(\hslash\omega_{\rm vm})\approx0$]. In the range of 200-600~K , $\alpha(T)$ (Fig.~\ref{fig:PTO}d) exhibits the classical linear-$T$ dependencies as widely reported in the literature~\cite{remeika1970growth,ikegami1971electromechanical,haun1989thermodynamic,tuttle1980ferroelectric,haun1987thermodynamic}. This behavior arises from $\langle\delta{P^2_{\rm th}}\rangle\propto{n_B(\hslash\omega_{\rm vm})}\propto{T}$ at 200-600~K (inset of Fig.~\ref{fig:PTO}d). Above $600$~K, the increased $\langle\delta{P^2_{\rm th}}\rangle$ approaches the reduced $P^2$, causing a nearly vanishing excitation gap $\Delta$ with rapidly enhanced fluctuation $\langle\delta{P^2_{\rm th}}\rangle$ and hence $\alpha(T)$ toward zero. 
  Thus, the phase transition occurs around the point where the magnitude of order parameter fluctuation exceeds the magnitude of the order parameter. The $\beta(T)$ remains negative around critical temperature (Fig.~\ref{fig:PTO}d), and the phase transition is therefore first-order.

Knowing the full temperature dependence of the polarization $P(T)$ and its thermal fluctuations allows us to directly predict a broad range of additional materials properties and device-relevant behaviors. For instance, for the rate of change of polarization, known as the pyroelectric coefficient $dP/dT$, one of the key functional coefficients central to sensor and energy-harvesting applications, our theoretically produced results show quantitative agreements with experimental measurements~\cite{remeika1970growth}, as seen in the inset of Fig.~\ref{fig:PTO}c. Furthermore, because the unit-cell volume change in tetragonal PbTiO$_3$ scales as $\delta V \propto P^2$~\cite{lines2001principles,Shirane1951}, showing the intimate coupling between ferroelectric order and structural distortion,  the theoretically predicted temperature dependence of the volume strain (Fig.~\ref{fig:PTO}e) for this thermomechanical response naturally emerges from the same $P^2(T)$ curve and again exhibits excellent quantitative agreement with experimental data~\cite{Shirane1951}.  Another important quantity governed by polarization fluctuations is the specific heat. The fluctuation contribution to the constant-pressure specific heat, $c_{p}=N_A\Omega_{\rm cell}\partial_T[\sum_{\bf q}\hslash\omega_{\rm vm}n_B(\hslash\omega_{\rm vm})]$ is expected to become significant near $T_c$ due to the rapid enhancement of thermal fluctuations as the excitation gap $\Delta$ collapses. As shown in Fig.~\ref{fig:PTO}f, the calculated $c_p(T)$ profile, including both the overall magnitude and the sharp anomaly at the transition, quantitatively reproduces the experimentally measured behavior~\cite{Yoshida1960}.   Importantly, all of these agreements are achieved using only the zero-temperature parameters obtained from independent low-temperature measurements, without adjusting any parameters at finite temperatures.

These results show that the developed framework enables one to use only the zero-temperature parameters to not only quantitatively capture the ferroelectric
properties at finite temperatures (phase evolution of order parameter with increasing temperature) in the entire range of the phase, including the classical criticality of the ferroelectric phase transition, but also be able to accurately reproduce a wide variety of experimentally measurable properties across a wide temperature range, including dielectric response, ferroelectric order, structural distortion, pyroelectric coefficient, and specific-heat anomaly.

\begin{figure}[htb]
  {\includegraphics[width=8.7cm]{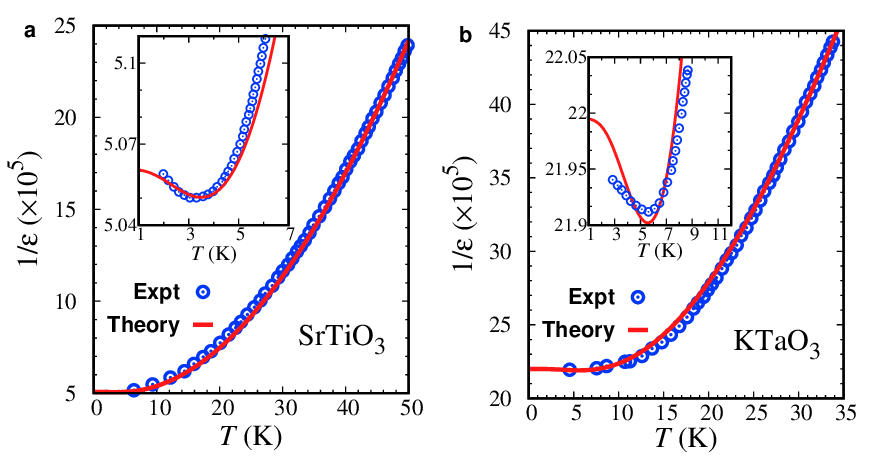}}
  \caption{Temperature dependence of inverse dielectric function in quantum paraelectric {\bf a}, SrTiO$_3$ and {\bf b}, KTaO$_3$. The solid curves are predicted values from the present theory, and  the dots come from the experimental measurement in Ref.~\cite{rowley2014ferroelectric}. The insets show the results at low-temperature limit.}    
\label{fig:STO}
\end{figure}

\section*{Application to incipient ferroelectricity}

We next consider the textbook examples of displacive quantum paraelectrics SrTiO$_3$ and KTaO$_3$~\cite{li2019terahertz,cheng2023terahertz,rowley2014ferroelectric,cowley1996phase}. 
 The structures of SrTiO$_3$ and KTaO$_3$ are found to be cubic perovskite  at room temperature.  While KTaO$_3$ remains as cubic down to low-temperature limit~\cite{singh1996stability},  SrTiO$_3$ undergoes a cubic-to-tetragonal antiferrodistortive transition at $105~$K~\cite{shirane1969lattice,vogt1995refined,cowley1996phase}, but its tetragonality of $1.00056$ is very small, causing essentially no change in the cell volume, thermal-expansion coefficient, or dielectric properties at the antiferrodistortive transition~\cite{beattie1971pressure,tao2016nonmonotonic}.
 Thus, we can directly use the model developed in the Model section [Eqs.~(\ref{Ham})-(\ref{theq})]. Existing measurements on SrTiO$_3$ and KTaO$_3$ show 
\begin{equation}
\lambda\approx0,
\end{equation}
yielding null renormalization $\beta(T)=b_0=b$ in our framework. We thus   only focus on the renormalization of $a$ to $\alpha(T)$.

In the so called quantum paraelectrics, there exists a strong competition between the quantum fluctuation and ferroelectric ordering~\cite{muller1979srti}, i.e., there exist the unstable phonon modes related to the ferroelectricity, but the zero-point vibrations of the lattice dynamics destroy the long-range ordering~\cite{li2019terahertz,cheng2023terahertz}.  In other words, one has a negative bare ground-state  parameter $a<0$ by first-principles calculations~\cite{verdi2023quantum}, but the renormalization from the zero-point oscillation of the collective vectorial polar mode yields  
 a nearly vanishing $a_0>0$ (an incipient ferroelectricity), i.e.,
 \begin{equation}
 a<0\xrightarrow{\text{zero-point renormalization}}a_0>0.
 \end{equation}
For SrTiO$_3$, the calculated $a\sim-70.15\times10^{-5}/\varepsilon_0$ from PBEsol functional~\cite{verdi2023quantum}, suggesting a ferroelectric ground state at the bare case. Nevertheless, after the self-consistent renormalization by the zero-point fluctuations, our calculation predicts a positive but nearly vanishing $a_{0}\approx3.21\times10^{-5}/\varepsilon_0$, close to the experimentally measured value  $a_0\approx5.06\times10^{-5}/\varepsilon_0$~\cite{rowley2014ferroelectric}. This description confirms the established understanding of the quantum paraelectric ground state of SrTiO$_3$ reported in the literature, and further confirms the crucial role of anharmonic contributions and zero-point lattice fluctuations in determining the $T=0$ parameters in the first-principles calculations of SrTiO$_3$,  as reported in Refs.~\cite{verdi2023quantum,he2020anharmonic}.

\begin{figure}[htb]
  {\includegraphics[width=8.7cm]{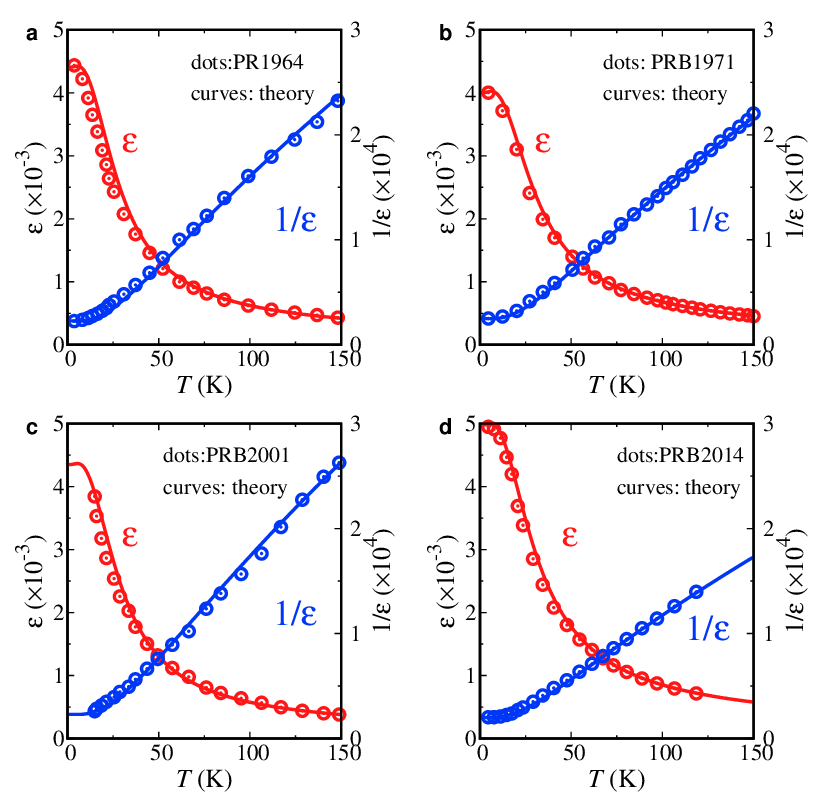}}
  \caption{{{Dielectric function and inverse dielectric function versus temperature in quantum paraelectric KTaO$_3$. In the figure,  the solid curve are predicted values from the present theory, and the experimental results (circles) in {\bf a}, {\bf b}, {\bf c} and {\bf d} come from Refs.~\cite{wemple1965some},~\cite{abel1971effect},~\cite{ang2001dielectric} and~\cite{aktas2014polar}.  The specific zero-temperature parameters, and their determination are addressed in Appendix. }} }    
\label{fig:KTO}
\end{figure}

\section*{Application to quantum paraelectrics SrTiO$_3$ and KTaO$_3$}

Further applying the renormalization by thermal fluctuations on the experimentally measured positive $a_0$~\cite{rowley2014ferroelectric},
 \begin{equation}
 a_0>0\xrightarrow{\text{thermal-fluctuation renormalization}}\alpha(T), 
 \end{equation}
we predict the dielectric properties of quantum paraelectrics SrTiO$_3$ and KTaO$_3$ at finite temperatures, and plot their inverse dielectric functions in Fig.~\ref{fig:STO}. The  thermal excitation of the collective vectorial polar mode in this case is expected to prohibit the ferroelectric ordering, driving system away from ferroelectricity.  {The used specific zero-temperature parameters $a_0$, $b_0$ and $g$, and their determination using several independent measurements~\cite{rowley2014ferroelectric,yamada1969neutron,fleury1968electric,yamada1969neutron}
are addressed in Appendix [See Table~SII for SrTiO$_3$ and Table~SIII for KTaO$_3$ in Appendix].} As seen from the figures, the predicted temperature dependencies of $1/\varepsilon(T)$ from the present theory almost completely quantitatively coincide with the experimental measurements, and this remarkable quantitative agreement  is achieved only using the zero-temperature parameters, determined from independent experimental measurements without adjusting parameters or $T$-dependent tuning.

More specifically, with nearly vanishing $a_0$ and $P^2(T)\equiv0$, the excitation gap of the collective vectorial polar mode $\Delta$ [Eq.~(\ref{gap})] nearly vanishes near $T=0$ and gradually increases with temperature due to the increased thermal fluctuation. This gap behavior in the bosonic thermal excitation leads to the $T$-square and linear-$T$ dependencies of $\langle{\delta{P^2_{\rm th}(T)}}\rangle$ at low and high temperatures, respectively. As a result, the inverse dielectric function exhibits a non-classical $T$-square dependence in a wide temperature range, e.g., $5$-$50$~K for SrTiO$_3$ (Fig.~\ref{fig:STO}a) and of $7$-$35$~K for KTaO$_3$ (Fig.~\ref{fig:STO}b).

Above this range, $1/\varepsilon(T)$ exhibits a classical linear-$T$  (i.e., Curie–Weiss) behavior.  The measurement in Ref.~\cite{rowley2014ferroelectric} did not extend to the linear-$T$ regime, while the linear-$T$ behavior of $1/\varepsilon(T)$ at high temperatures has been reported in the literature in both SrTiO$_3$~\cite{yang2022epitaxial} and KTaO$_3$~\cite{wemple1965some,abel1971effect,ang2001dielectric,aktas2014polar}. To validate our framework, we utilize several experimental reports~\cite{wemple1965some,abel1971effect,ang2001dielectric,aktas2014polar} in KTaO$_3$ for comparison, and plot the dielectric functions in Fig.~\ref{fig:KTO}. For each experiment, the predicted temperature dependencies of the dielectric properties from our theory quantitatively coincide with the measured ones not only in the non-classical $T$-square regime at relatively low temperatures but also in the classical linear-$T$ regime at high temperatures.

At low-temperature limit (1-7~K), due to the minimal thermal fluctuation, the inverse dielectric function is expected to be insensitive to temperature variation and saturates to a plateau. However, with the decrease in temperature towards zero, experiments have observed a small upturn of $1/\varepsilon(T)$ below a few kelvin~\cite{rowley2014ferroelectric}. This anomalous upturn is suggested to be relevant to the coupling of the polarization field with gapless acoustic phonons through electrostrictive effect~\cite{palova2009quantum,khmelnitskii1973phase,rowley2014ferroelectric,fujishita2016quantum}. Following the same way, here we include this effect/correction  in the renormalization via the correction term $\delta\alpha(T)$ in Eq.~(\ref{alpha}), by adding thermal excitation of the acoustic phonons (see Appendix). The electrostrictive coefficient and parameters of the acoustic phonons used in our calculation are determined by independent measurements. The predicted results shown in the insets of Fig.~\ref{fig:STO}a for SrTiO$_3$ and of Fig.~\ref{fig:STO}b for KTaO$_3$ also exhibit a quantitative agreement with the experimental measurements.

\section*{Application to ferroelectric strained SrTiO$_3$}

With rapid advances in materials synthesis, accessing hidden ferroelectric phases through strain engineering has become particularly appealing, as it offers a clean and reversible tuning knob that does not introduce disorder or extrinsic carriers,  yet stabilizes the ferroelectric order in systems that remain paraelectric in their  bulk  form~\cite{haun1987thermodynamic,lines2001principles,haun1989thermodynamic}.  A  recent experiment on SrTiO$_3$ membranes~\cite{li2025classical} reported that the strain-induced ferroelectric transition temperature follows a scaling relation that deviates from the classical Landau-type prediction, $T_c \propto (s-s_{\rm cp})^{0.5}$, where $s_{\rm cp}$ is the critical strain for the zero-temperature ferroelectric transition. This deviation was  interpreted as a signature of the growing influence of the quantum fluctuations near  quantum critical regime, which are believed to modify the critical exponent of the power-law scaling from its classical value of 0.5 to approximately 0.606 in the strained SrTiO$_3$. However, attributing the deviation of finite-temperature critical exponents solely to quantum fluctuations may be misleading. The core issue lies in the applicability of the classical Landau theory, whereas in the present context, particularly in the low-temperature regime relevant to quantum criticality, a more natural and physically consistent explanation should involve properly accounting for thermal fluctuations within the thermal-bosonic-excitation framework, rather than invoking the zero-point (quantum) fluctuations.

Our framework  can be naturally applied to describe this strain-induced ferroelectricity, enabling quantitative predictions of their dielectric and ferroelectric properties at finite temperatures and under nonzero external strain in the entire phase regime, using only strain-free, zero-temperature parameters. Specifically, focusing on SrTiO$_3$, we incorporate the electrostrictive coupling~\cite{khmelnitskii1973phase,haun1987thermodynamic,fujishita2016quantum,schimizu1997effect}
\begin{equation}
a_0\rightarrow{a_0}-C_0s,
\end{equation}
into our framework, with $C_0$ the coupling constant and $s$ the external strain. Then, self-consistently solving the renormalization solely from  thermal fluctuations yields a temperature- and strain-dependent $\alpha(T,s)$,
 \begin{equation}
 a_0-C_0s\xrightarrow{\text{thermal-fluctuation renormalization}}\alpha(T,s).  
 \end{equation}
 This allows us to determine the critical behavior, such as the strain dependence of the ferroelectric transition temperature $T_c(s)$, by solving the critical condition $\alpha(T_c, s) = 0$.

\begin{figure}[htb]
  {\includegraphics[width=8.8cm]{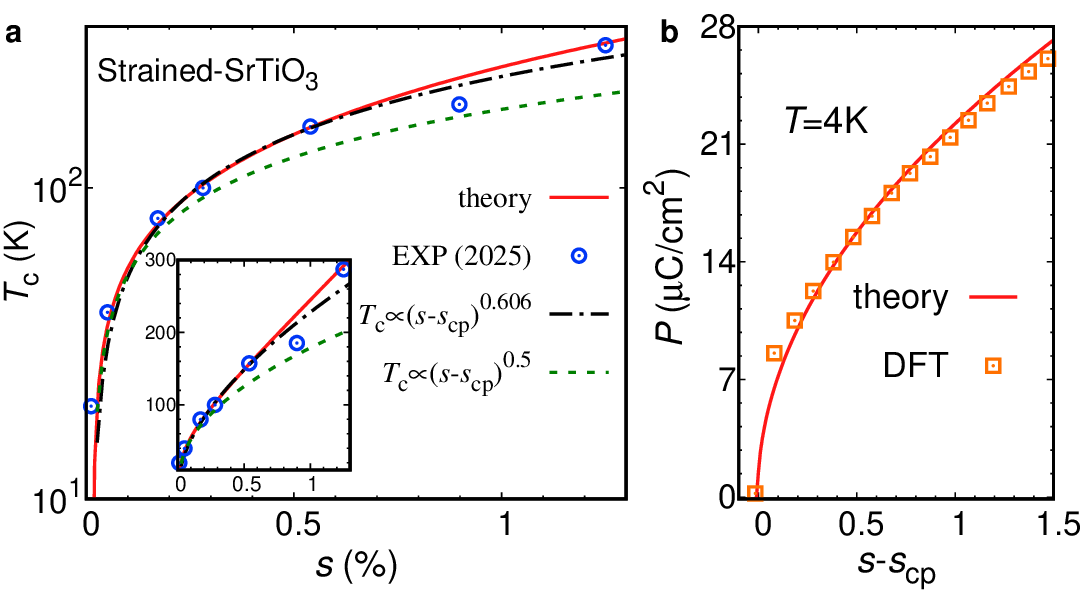}}
  \caption{(a) Ferroelectric transition temperature as a function of externally applied strain. The solid curve represent calculated results from the present  renormalization theory,  whereas the data points are experimental measurements from Ref.~\cite{li2025classical}.  The chain curve denotes an empirical power-law relation $T_c \propto (s-s_{\rm cp})^{0.606}$ proposed by experiments in Ref.~\cite{li2025classical}, and the dashed curve comes from the prediction by using Landau theory~\cite{li2025classical}, which predicts a scaling $T_c \propto (s-s_{\rm cp})^{0.5}$ when assuming $\alpha(T)\sim{T^2}$ (as 
  suggested by experiment~\cite{rowley2014ferroelectric}). The inset shows the same data,  but plotted with a linear vertical axis instead of a logarithmic one. (b) Calculated polarization as a function of strain
at fixed low temperature from our theory and Density functional theory. DFT results are
taken from Ref.~\cite{xu2020strain}.} 
\label{StrainedSTO}
\end{figure}

For the experimentally measured SrTiO$_3$ membrane samples~\cite{li2025classical,xu2020strain}, the strain-free, zero-temperature parameters are not available from existing studies. The electrostrictive coupling constant $C_0$ can be determined from the experimentally observed quantum critical point~\cite{li2025classical}, characterized by a critical strain $s_{\rm cp}$, using the theoretical condition $a_0-C_0s_{\rm cp}=0$ for the zero-temperature ferroelectric transition{, i.e.,  $C_0=a_0/s_{\rm cp}$ with $s_{\rm cp}=0.019\%$~\cite{li2025classical}}. All other model (strain-free, zero-temperature) parameters are taken to be the same as those of the bulk material, which serves as a reasonable approximation given the high crystalline quality and weak strain-relaxation effects in the membrane sample.

The predicted results for strained SrTiO$_3$ membrane are shown in Fig.~\ref{StrainedSTO}.  Remarkably, as shown in Fig.~\ref{StrainedSTO}a, without adjusting or fitting any parameters, the predicted strain dependence of the ferroelectric transition temperature $T_c$ (solid curve) directly reproduces the experimental measurements (circles) with excellent quantitative agreement over a wide range of applied strain. This prediction outperforms the empirical laws $T_c \propto (s-s_{\rm cp})^{0.606}$ (chain curve) proposed by experiments in Ref.~\cite{li2025classical} as well as the Landau-theory prediction  $T_c \propto (s-s_{\rm cp})^{0.5}$ (dashed curve)~\cite{li2025classical} obtained by assuming $\alpha(T)\sim T^2$ (as suggested by experiment~\cite{rowley2014ferroelectric}). We emphasize again that 
this remarkable  quantitative agreement is directly achieved  using only strain-free, zero-temperature parameters without any fitting.  Consequently, our results clearly demonstrate that once the bosonic nature of the polar modes (polarization fluctuations) is properly incorporated within the thermal ensemble, the full experimental strain dependence of $T_c(s)$ can be quantitatively reproduced without invoking quantum fluctuations of the order-parameter field. From the viewpoint of quantum statistical mechanics~\cite{abrikosov2012methods}, in the low-temperature regime where the bosonic nature of thermal excitations dominates, the system no longer follows classical critical scaling; in this regime, both the classical Landau exponent and its empirically modified variants become inappropriate, signaling a breakdown of classical power-law criticality. This demonstrates the necessity of our quantum-statistic framework, which provides a unified and self-consistent description of ferroelectric behaviors
across the entire phase 
 that are inaccessible to phenomenological Landau-Ginzburg  approaches.  

As shown in Fig.~\ref{StrainedSTO}b, in the low-temperature limit, our predicted strain dependence of the polarization also shows excellent quantitative agreement with first-principles (DFT) results, further validating the microscopic fidelity of the present framework. This agreement indicates that the underlying polarization formation mechanism and its coupling to strain are captured at a level fully consistent with ab-initio calculations, yet achieved at dramatically reduced computational cost, while still retaining the ability to accurately and quantitatively extend to finite-temperature and even nonequilibrium regimes. \\

\section*{Discussion}

We develop a self-consistent computational framework for finite-temperature properties and phase-transitions of the displacive ferroelectricity based on the quantum statistic theory, linking the
condensed ground state to displacive ferroelectric phase transitions. It  enables one to use only the
ground-state parameters to predict the dielectric/ferroelectric
properties at finite temperatures in the entire range of the phase, including the  criticality of the phase. If one uses first-principles calculations  to determine these ground-state parameters as input, our framework bridges them to finite-temperature behavior across entire temperature range and to critical phenomena. In cases where first-principles calculations may not fully capture essential effects, experimentally measured zero-temperature parameters can be used to reliably apply the approach. 

We have demonstrated its applications to three types of representative materials systems. The framework therefore offers a useful theoretical tool for understanding and predicting many of the thermodynamic and non-equilibrium behaviors of displacive ferroelectricity. For example, a fully self-consistent computation of the underlying excitation spectrum naturally allows us to quantitatively calculate the thermal properties (e.g., specific heat and thermal conductivity~\cite{3y1m-66s1}) as well as the non-equilibrium dynamics (e.g., THz response in quantum paraelectrics~\cite{li2019terahertz,cheng2023terahertz,74d5-4hsw}) that rely on quasiparticle description. It can also be applied to describe the multi-structural phase transitions in BaTiO$_3$, using solely on one set of ground-state parameters.  From a more fundamental viewpoint, this theory is a general phase-transition theory of the BEC, and it may be relevant to understanding a broad range of the phase transitions in various bosonic systems, such as systems exhibiting antiferrodistortive transitions, and beyond.

{The accuracy of the theoretical predictions depends on the reliability of the low-temperature experimental inputs. On the other hand, within our framework, the dominant contribution is captured by the collective polar modes (i.e., soft-mode fluctuations), while additional subleading corrections, such as the perturbative coupling of the polarization field to other phonon branches (e.g., acoustic phonons) are included at the perturbative level. In the case of experimentally well-studied systems such as PbTiO$_3$, SrTiO$_3$, and KTaO$_3$, the currently available experimental data in the low-temperature regime are well established, and the corresponding subleading corrections are quantitatively small (as shown in the insets of Fig.~\ref{fig:STO}). As a result, the overall uncertainty is limited and does not affect the main conclusions of the present work.}

{The key differences between the present framework and the conventional Landau-Ginzburg approach manifest in two main aspects in the present study. On one hand, taking PbTiO$_3$ as an example, the Landau theory typically assumes prior knowledge of the transition temperature $T_c=763~K$ and the first-order nature of the transition, and introduces a phenomenological expansion such as $\alpha(T)=a_0(1-T/T_c)$ in the vicinity of $T_c$.In contrast, within our framework, both the transition temperature and the first-order character of the transition emerge naturally from the underlying quantum-statistical and many-body treatment, without being imposed a prior. On the other hand, for quantum paraelectrics SrTiO$_3$ and KTaO$_3$, the conventional Landau approach typically predicts a classical linear-$T$ (i.e., Curie-Weiss) behavior of $1/\varepsilon(T)$. However, as shown in Figs.~\ref{fig:STO} and~\ref{fig:KTO}, such linear behavior is only recovered at high temperatures, while at low temperatures the system exhibits a non-classical $T^2$ dependence. Beyond relying solely on ground-state parameters, the present framework also offers several key advantages. It provides a transparent physical interpretation in terms of collective soft-mode fluctuations, allowing a direct connection between microscopic lattice dynamics and macroscopic polarization. It also enables a unified description of both quantum and thermal fluctuations within a self-consistent statistical framework, which is essential for capturing thermodynamic behavior. The approach  naturally incorporates strain effects through the coupling between polarization and lattice degrees of freedom, and is broadly applicable to different classes of displacive ferroelectrics, providing a general route for quantitatively linking microscopic properties to finite-temperature functional responses.}

{\bf Implementation details}
For both ferroelectric PbTiO$_3$ and quantum paraelectrics SrTiO$_3$ and KTaO$_3$, in evaluating the bosonic excitation spectrum of the collective vector mode, a wave-vector cutoff $q_{c,\mathrm{FE}}$ is introduced to regularize the momentum integrals. The key parameters of the model, including $a_0$ and $b_0$ for SrTiO$_3$ and KTaO$_3$, as well as $a_i$, $a_{ij}$, and $a_{ijk}$ for PbTiO$_3$, together with $g$ and $q_{c,\mathrm{FE}}$, which govern the dominant ferroelectric and paraelectric properties, are consistently determined from low-temperature experimental inputs (approximately 4~K). For smaller correction terms (e.g., those responsible for the upturn features shown in the insets of Fig.~3), certain parameters, such as those associated with acoustic phonons and their coupling to the polarization field, are not directly accessible from low-temperature experiments. In these cases, we estimate them based on a combination of first-principles calculations and available experimental data. These corrections have only minor quantitative effects on the main results and do not affect the predictive capability of the model for the dominant dielectric and ferroelectric behavior. The specific parameter sets used for PbTiO$_3$, SrTiO$_3$, and KTaO$_3$, as well as their determination, are summarized in Tables~SI-SIII of the Appendix. The self-consistent equations are solved iteratively until a relative convergence criterion of $10^{-6}$ is reached.

{\bf Acknowledgments}
  This work is supported by the US Department of Energy, Office of Science, Basic Energy Sciences, under Award Number DE-SC0020145 as part of Computational Materials Sciences Program.  F.Y. and L.Q.C. also appreciate the generous support from the Donald W. Hamer Foundation through a Hamer Professorship at Penn State.

\begin{widetext}
\begin{appendix}
\setcounter{equation}{0}
\counterwithout{equation}{section}
\renewcommand{\thesection}{Appendix~S\Roman{section}} 
\renewcommand{\theequation}{S\arabic{equation}}
\renewcommand{\thetable}{S\Roman{table}}
\renewcommand{\thefigure}{S\Roman{figure}}
\renewcommand{\thesubsection}{\Alph{subsection}}

\section{Detailed derivation of the model}
\label{III}
Based on the ground-state mean-field Hamiltonian in the main text, we start with the action of the polarization field,
\begin{equation}
\mathcal{S}=\int{dt}d{\bf x}\Big\{\frac{m_p}{2}(\partial_t{\hat P})^2-\Big[\frac{g}{2}({\bf \nabla}{\hat P})^2+\frac{a}{2}{\hat P}^2+\frac{b}{4}{\hat P}^4+\frac{\lambda}{6}{\hat P}^6\Big]\Big\},
\end{equation}  
which with the polarization field ${\bf \hat P}={\bf P}+\delta{\bf P}$ can be re-written as
\begin{eqnarray}
\mathcal{S}&=&\int{dtd{\bf x}}\Big\{\!\frac{m_p}{2}(\partial_t\delta{P})^2-\big[\frac{g}{2}({\bf \nabla}\delta{P})^2+\frac{a}{2}{\delta}P^2+\frac{b}{4}{\delta}P^4+\frac{\lambda}{6}{\delta}P^6\big]-\big[\frac{b}{2}\big(\frac{2}{d_p}+1\big){P}^2\delta{P}^2+\frac{\lambda}{2}\big(\frac{4}{d_p}+1\big)({P}^4\delta{P}^2+{P}^2\delta{P}^4)\big]\nonumber\\
&&\mbox{}-\big(\frac{a}{2}P^2+\frac{b}{4}P^4+\frac{\lambda}{6}P^6\big)\Big\},\label{LAG}
\end{eqnarray}
where we have taken the angular average of the fluctuations by neglecting their odd orders and considering isotropic fluctuations. Then,  the thermally averaged contributions purely from the long-range ordering in Eq.~(\ref{LAG}) give rise to the free energy. In addition, with this action, the Euler-Lagrange equation of motion with respect to the polarization fluctuation $\delta{\bf P}$ leads to
\begin{equation}
\big[m_p\partial_t^2+\gamma\partial_t-g\nabla^2+a+b(2/d_p+1){P}^2+\lambda(4/d_p+1){P}^4+b\delta{P}^2+\lambda\delta{P}^4+2\lambda(4/d_p+1){P}^2\delta{P}^2\big]\delta{\bf P}={\bf E}_{\rm th}(t,{\bf R}).  \label{PFEq}
\end{equation}
Here, we have introduced a damping rate $\gamma=0^+$ and a thermal field ${\bf E}_{\rm th}(t,{\bf R})$ that  obeys the fluctuation-dissipation theorem~\cite{landau1981statistical}:
\begin{equation}
\langle{\bf E}_{\rm th}(\omega,{\bf q}){\bf E}_{\rm th}(\omega',{\bf q}')\rangle=\frac{\gamma\hslash\omega(2\pi)^4\delta({\bf q}-{\bf q'})\delta(\omega-\omega')}{\tanh[\hslash\omega/(2k_BT)]},  
\end{equation}
with the four-momenutm $p=(\omega,{\bf q})$ being the Fourier space of the space-time coordinate $x=(t,{\bf R})$. Then, under the mean-field approximation, combined with the assumption of isotropic fluctuations, yielding:
\begin{eqnarray}
 \delta P^2 \delta {\bf P} \rightarrow \Big(\frac{2}{d_p}+1\Big)\langle\delta P^2 \rangle \delta {\bf P},
\end{eqnarray} 
or 
\begin{eqnarray}
 \delta P^2 \delta {\bf P}&=&\sum_{i}\delta P^2 \delta {P}_i{\bf e}_i=\sum_{i}\delta P^2_i \delta {P}_i{\bf e}_i+\sum_{i{\ne}j}\delta P^2_j \delta {P}_i{\bf e}_i\rightarrow3\sum_{i}\langle\delta P^2_i\rangle \delta {P}_i{\bf e}_i+\sum_{i{\ne}j}\langle\delta P^2_j\rangle \delta {P}_i{\bf e}_i\nonumber\\
 &=&2\sum_{i}\langle\delta P^2_i\rangle \delta {P}_i{\bf e}_i+\sum_{i}\langle\delta P^2\rangle \delta {P}_i{\bf e}_i=\frac{2}{d_p}\sum_{i}\langle\delta P^2\rangle \delta {P}_i{\bf e}_i+\sum_{i}\langle\delta P^2\rangle \delta {P}_i{\bf e}_i=\Big(\frac{2}{d_p}+1\Big)\langle\delta P^2 \rangle \delta {\bf P},
\end{eqnarray}
we have
\begin{equation}
\big[m_p\partial_t^2+\gamma\partial_t-g\nabla^2+m_p\Delta^2\big(a,b,P^2,\langle\delta{P}^2\rangle\big)\big]\delta{\bf P}={\bf E}_{\rm th}(t,{\bf R}). 
\end{equation}
with the excitation gap:
\begin{equation}
\label{sigap}
\Delta\!=\!\sqrt{\frac{a\!+\!b(2/d_p\!+\!1){P}^2\!+\!\lambda(4/d_p\!+\!1){P}^4\!+\!b(2/d_p\!+\!1)\langle\delta{P}^2\rangle\!+\!3\lambda(4/d_p\!+\!1)\langle\delta{P}^2\rangle^2\!+\!2\lambda(4/d_p\!+\!1)(2/d_p\!+\!1){P}^2\langle\delta{P}^2\rangle}{m_p}},~~~
\end{equation}
and then, one can directly solve the thermally averaged polarization fluctuation:
\begin{equation}\label{siPF}
  \langle{\delta{P}^2}\rangle=\int\frac{\hslash}{2m_p}\frac{\coth\big[\hslash\omega_{vm}\big(q,a,b,P^2,\langle\delta{P^2}\rangle\big)/(2k_BT)\big]}{\omega_{vm}\big(q,a,b,P^2,\langle\delta{P^2}\rangle\big)}\frac{d{\bf q}}{(2\pi)^3},  
\end{equation}
where the energy spectrum of the collective vectorial polar mode {\small $\hslash\omega_{vm}\big(q,a,b,P^2,\langle\delta{P^2}\rangle\big)=\hslash\sqrt{\Delta^2\big(a,b,P^2,\langle\delta{P^2}\rangle\big)+gq^2/m_p}$} with the excitation gap $\Delta\big(a,b,P^2,\langle\delta{P^2}\rangle\big)$ given in Eq.~(\ref{sigap}). The polarization fluctuation here consists of the contributions from the zero-point oscillation and thermal excitation in the Bosonic excitation of the collective vectorial polar mode. 

In renormalization theory, with a vanishing thermally averaged $\langle\delta{\bf P}\rangle$ but a nonzero thermally averaged $\langle{\delta{P^2}}\rangle$, from Eq.~(\ref{LAG}),  one can extract the contributions purely from the long-range ordering:
\begin{equation}
\mathcal{S}_{P}=\int{d{\bf x}}\Big[\frac{\bar a}{2}P^2+\frac{\bar b}{4}P^4+\frac{\lambda}{6}P^6\Big],
\end{equation}
where
\begin{eqnarray}
 {\bar a}&=&a\!+\!b(2/d_p\!+\!1)\langle\delta{P}^2\rangle\!+\!3\lambda(4/d_p\!+\!1)\big(\langle\delta{P}^2\rangle\big)^2, ~~~\label{sialpha}\\
  {\bar b}&=&b+2\lambda(4/d_p\!+\!1)\langle\delta{P}^2\rangle, ~\label{sibeta}
\end{eqnarray}
with $d_p$ being the dimension of the vector space of $\delta{\bf P}$.

We next introduce the proposed renormalization process. 

{\sl Renormalization by zero-point fluctuations.---}At zero temperature, one can summarize Eqs.~(\ref{sialpha}) and (\ref{sibeta}) as well as Eqs.~(\ref{siPF}) and~(\ref{sigap}), and directly find the self-consistent renormalization processes by the zero-point fluctuations:
\begin{eqnarray}
a_0&=&a+b(2/d_p\!+\!1)\langle\delta{P}^2_{\rm zo}\rangle+3\lambda(4/d_p\!+\!1)\big(\langle\delta{P}^2_{\rm zo}\rangle\big)^2+\delta{a}, ~~~\\
b_0&=&b+2\lambda(4/d_p\!+\!1)\langle\delta{P}^2_{\rm zo}\rangle, 
\end{eqnarray}  
where the zero-point fluctuations:
\begin{equation}
 \langle{\delta{P}^2_{\rm zo}}\rangle=\int\frac{\hslash}{2m_p}\frac{1}{\omega_{vm}(q,a,b,P_0^2,\langle\delta{P_{\rm zo}^2}\rangle)}\frac{d{\bf q}}{(2\pi)^3},  
\end{equation}
and the zero-temperature long-range ordered polarization: $P_0^2=\big(-b_0+\sqrt{b_0^2-4\lambda{a_0}}\big)/(2\lambda)~$ for $a_0<0$ and $P_0^2=0$ for $a_0>0$.

{\sl Renormalization by thermal fluctuations.---}After the renormalization processes by the zero-point fluctuations, the self-consistent renormalization processes by the thermal fluctuations at finite temperatures are given by
\begin{eqnarray}
  \alpha&=&a_0+b_0(2/d_p\!+\!1)\langle\delta{P}^2_{\rm th}(T)\rangle+3\lambda(4/d_p\!+\!1)\big(\langle\delta{P}^2_{\rm th}(T)\rangle\big)^2+\delta{\alpha}(T),\label{aaaa}~~~\\
  \beta&=&b_0+2\lambda(4/d_p\!+\!1)\langle\delta{P}^2_{\rm th}(T)\rangle, 
\end{eqnarray}
where the thermal fluctuations are determined by substracting the zero-temperature part from Eq.~(\ref{siPF}) and are written as 
\begin{equation}
\langle{\delta{P}^2_{\rm th}}\rangle=\int\frac{\hslash}{2m_p}\Bigg\{\frac{\coth\big[\hslash\omega_{vm}\big(q,a_0,b_0,P^2,\langle\delta{P_{\rm th}^2}\rangle\big)/(2k_BT)\big]}{\omega_{vm}\big(q,a_0,b_0,P^2,\langle\delta{P_{\rm th}^2}\rangle\big)}-\frac{1}{\omega_{vm}\big(q,a_0,b_0,P_0^2,0\big)}\Bigg\}\frac{d{\bf q}}{(2\pi)^3},   
\end{equation}  
and finite-temperature long-range ordered polarization: $P^2=\big(-\beta+\sqrt{\beta^2-4\lambda\alpha}\big)/(2\lambda)$ for $\alpha<0$ and $P^2=0$ for $\alpha>0$. 

 We have introduced a correction/perturbation term, $\delta{a}$ and $\delta\alpha(T)$, which can account for the perturbative coupling of the polarization field to other phonon branches  (e.g., acoustic phonons via  electrostrictive~\cite{palova2009quantum,khmelnitskii1973phase,rowley2014ferroelectric,fujishita2016quantum} or flexoelectric~\cite{PhysRevLett.131.046801} coupling) or take account into the self-consistent structural instabilities around transition due to significant thermal excitation of the collective vectorial polar mode. 

\section{Derivation of the model within  fundamental path-integral approach in Matsubara representation}

The present work assumes that after the structural/ferroelectric phase transition in displacive ferroelectrics, the  collective vectorial polar mode   emerges and determines the dielectric/ferroelectric properties of the system, as a direct and fundamental consequence of the condensation of the imaginary-frequency phonons. These excitations emerge as a result of the renormalized dynamics of the system as a whole after the emergence of the electric polarization, rather than being direct representations of individual original phonons. Our theory is  concerned with the low-lying excitations associated with the long-range fluctuations, which should dominate the fluctuations~\cite{patashinskii1979fluctuation,strukov2012ferroelectric} over a large temperature regime above zero as the short-range/local excitation typically has a higher excitation energy. 

For checking the self-consistency of our proposed microscopic thermodynamic theory, below we present a separate framework to derive the same theory using the fundamental path-integral approach in the Matsubara (imaginary-time) representation. This derivation leads to the same theoretical description as the one in the Methods section of the main text, which is derived by using the fluctuation dissipation theorem via introducing a thermal field. The basic idea is to derive the effective action (i.e., partition function) of the global polarization ${\bf P}(T)$ within the path-integral approach, by self-consistently integrating out the fluctuation $\delta{\bf P}$ in Matsubara representation from initial action. 

Specifically, with the Hamiltonian  of the polarization field, the action in the Matsubara representation is written as
\begin{equation}\label{action}
  S\!=\!\int_0^{\hslash/(k_BT)}d{\tau}d{\bf q}\Big[\frac{m_p}{2}(\partial_{\tau}\delta{P})^2\!+\frac{gq^2}{2}\delta{P}^2\!+\!\frac{a}{2}{\delta}P^2\!+\!\frac{b}{4}{\delta}P^4\!+\!\frac{\lambda}{6}{\delta}P^6\!+\!\frac{b}{2}\big(\frac{2}{d_p}\!+\!1\big){P}^2\delta{P}^2\!+\!\frac{\lambda}{2}\big(\frac{4}{d_p}\!+\!1\big)({P}^4\delta{P}^2\!+\!{P}^2\delta{P}^4)\Big]+S_G,~~~~
\end{equation}
with $S_G=\frac{a}{2}P^2+\frac{b}{4}P^4+\frac{\lambda}{6}P^6$. To obtain the effective action (partial function) of the long-range ordered polarization ${P^2}$, one can perform the standard integration over the bosonic field of the polarization fluctuation $\delta{P}$ within the path-integral formalism. However, to tackle the phase transition, it requires formulating the infinite-order perturbation expansions of the self-energy correction due to the coupling between the polarization fluctuation and long-range ordered polarization, or picking up the appropriate diagrammatic representation in a self-consistent way, making it a challenging problem. Here we present two self-consistent methods similar to the treatments of superconductors~\cite{yang2024arXiv,yang2021theory,yang2022impurity,sun2020collective,abrikosov2012methods}. 

{\sl Generating functional methods.---}Within the path-integral formalism, using the action in Eq.~(\ref{action}), the thermal average of the polarization fluctuation can be written as~\cite{peskin2018introduction,abrikosov2012methods}
\begin{align}
  \langle{\delta{P}^2}\rangle=&\int\frac{d{\bf q}}{(2\pi)^3}\Big\langle\Big|\delta{P}(\tau,{\bf q})\delta{P}(\tau,-{\bf q})e^{-\frac{S}{\hslash}}\Big|\Big\rangle=\int\frac{d{\bf q}}{(2\pi)^3}\int\frac{D\delta{P}}{\mathcal{Z}_0}\Big[\delta{P}(\tau,{\bf q})\delta{P}(\tau,-{\bf q})e^{-\frac{S}{\hslash}}\Big]\nonumber\\
  =&\int\!\!\frac{d{\bf q}}{(2\pi)^3}\int\!\!\frac{D\delta{P}}{\mathcal{Z}_0}~\delta_{J_{\bf q}}\delta_{J_{-{\bf q}}}\big\{e^{-\frac{S}{\hslash}-\int_0^{\frac{\hslash}{k_BT}}d\tau{d}{\bf q'}[J_{\bf q'}\delta{P}(\tau,{\bf q'})]}\big\}\big|_{J=0}\nonumber\\
  =&\!\int\!\!\frac{d{\bf q}}{(2\pi)^3}\int\!\!\frac{D\delta{P}}{\mathcal{Z}_0}e^{-\frac{S_G}{\hslash}}\delta^2_{J_{\bf q}}\Big\{e^{-\int_0^{\frac{\hslash}{k_BT}}d\tau{d}{\bf q'}\big\{\frac{1}{2\hslash}\delta{P}(\tau,{\bf q'})\big[m_p\omega^2_{vm}\big(q',a,b,P^2,\langle\delta{P^2}\rangle\big)-m_p\partial_{\tau}^2\big]\delta{P}(\tau,{\bf q'})+J_{\bf q'}\delta{P}(\tau,{\bf q'})\big\}}\Big\}\Big|_{J=0}\nonumber\\
  =&\int\frac{d{\bf q}}{(2\pi)^3}~\delta_{J_{\bf q}}^2\exp\bigg\{\int_0^{\hslash/(k_BT)}d\tau{d}{\bf q'}\Big[\frac{1}{2}J_{\bf q'}\frac{\hslash}{m_p\omega^2_{vm}\big(q',a,b,P^2,\langle\delta{P^2}\rangle\big)-m_p\partial_{\tau}^2}J_{\bf q'}\Big]\bigg\}\Big|_{J=J^*=0}\nonumber\\
  =&-\!\!\int\!\!\frac{d{\bf q}}{(2\pi)^3}\frac{k_BT}{\hslash}\sum_{n}\frac{\hslash/m_p}{(i\omega_n)^2\!-\!\omega^2_{vm}\big(q,a,b,P^2,\langle\delta{P^2}\rangle\big)}=\int\frac{d{\bf q}}{(2\pi)^3}\frac{\hslash}{m_p}\frac{\coth\big[\hslash\omega_{vm}\big(q,a,b,P^2,\langle\delta{P^2}\rangle\big)/(2k_BT)\big]}{2\omega_{vm}\big(q,a,b,P^2,\langle\delta{P^2}\rangle\big)},\label{pppp}~~~~~~
\end{align}
where the energy spectrum of the collective vector mode 
\begin{equation}
\hslash\omega_{vm}\big(q,a,b,P^2,\langle\delta{P^2}\rangle\big)=\hslash\sqrt{\Delta^2\big(a,b,P^2,\langle\delta{P^2}\rangle\big)+gq^2/m_p}
\end{equation}
and the excitation gap:
\begin{equation}
\Delta\!=\!\sqrt{\frac{a\!+\!b(2/d_p\!+\!1){P}^2\!+\!\lambda(4/d_p\!+\!1){P}^4\!+\!b(2/d_p\!+\!1)\langle\delta{P}^2\rangle\!+\!3\lambda(4/d_p\!+\!1)\langle\delta{P}^2\rangle^2\!+\!2\lambda(4/d_p\!+\!1)(2/d_p\!+\!1){P}^2\langle\delta{P}^2\rangle}{m_p}},~~~
\end{equation}
Here, we have utilized the mean-field approximation to obtain the thermally averaged $\omega^2_{vm}$. Additionally, $\omega_n=2n\pi{k_BT}/\hslash$ represents the bosonic Matsubara frequencies; $J_{\bf q}$ denotes the generating functional and $\delta{J_{\bf q}}$ stands for the functional derivative~\cite{peskin2018introduction,yang2020theory}; $\mathcal{Z}_0=\langle|e^{-S}|\rangle$ is the normalization factor. Then, substituting Eq.~(\ref{pppp}) to the action in Eq.~(\ref{action}) and extracting the thermally averaged contributions from $P^2$ afterwards, one finds the partial function (i.e., free energy) of the long-range ordered polarization.

{\sl Self-consistent Green function methods.---}We re-write the action in Eq.~(\ref{action}) as
\begin{equation}
  S\!=\!\int_0^{\hslash/(k_BT)}\!\!d{\tau}d{\bf q}\frac{1}{2}\delta{P}(\tau,{\bf q})D^{-1}_{vm}\big(\partial_{\tau},q\big)\delta{P}(\tau,{\bf q})+S_G,
\end{equation}
where the inverse Green function is defined as~\cite{peskin2018introduction,abrikosov2012methods}
\begin{equation}
  D^{-1}_{vm}\big(\partial_{\tau},q\big)=gq^2\!+\!a\!+\!{b}{\delta}P^2/2\!+\!{\lambda}{\delta}P^4/3\!+\!b\big({2}/{d_p}\!+\!1\big){P}^2\!+\!{\lambda}\big({4}/{d_p}\!+\!1\big)({P}^4\!+\!{P}^2\delta{P}^2)-m_p\partial_{\tau}^2.
\end{equation}
It is noted that here we have kept the coupling between the polarization fluctuation and long-range ordered polarization in the Green function rather than in the self-energy as in the conventional way. Then, performing the integration over the bosonic field of the fluctuation within the path-integral formalism, one finds the effective action:
\begin{equation}
  S_{\rm eff}=\frac{1}{2}\hslash{\rm \bar{T}r}\ln\big[{D^{-1}_{vm}(\partial_{\tau},q\big)}\big]+S_G.  
\end{equation}
Through the variation with respect to $P$, the equation to determine the long-range ordered polarization reads
\begin{equation}
 \frac{1}{2}\hslash{\rm \bar{T}r}\Big\{\partial_{P}\big[{D^{-1}_{vm}\big(\partial_{\tau},q\big)}\big]D_{vm}(\partial_{\tau},q)\Big\}+{\partial_{P}}S_G=0.  
\end{equation}
Imposing the mean-field approximation on the thermal average, the above equation becomes
\begin{eqnarray}
  && \left\langle\hslash{\rm \bar{T}r}\Big\{\frac{b\big({2}/{d_p}\!+\!1\big){P}\!+\!{\lambda}\big({4}/{d_p}\!+\!1\big)(2{P}^3\!+\!{P}\delta{P}^2)}{gq^2\!+\!a\!+\!{b}{\delta}P^2/2\!+\!{\lambda}{\delta}P^4/3\!+\!b\big({2}/{d_p}\!+\!1\big){P}^2\!+\!{\lambda}\big({4}/{d_p}\!+\!1\big)({P}^4\!+\!{P}^2\delta{P}^2)-m_p\partial_{\tau}^2}\Big\}\right\rangle\!+\!{\partial_{P}}S_G\!=\!0\nonumber\\
  &&\Rightarrow\!{\rm \bar{T}r}\Big\{\frac{\hslash[{b\big({2}/{d_p}\!+\!1\big){P}\!+\!{\lambda}\big({4}/{d_p}\!+\!1\big)(2{P}^3\!+\!3{P}\langle\delta{P}^2\rangle)}\langle\delta{P}^2\rangle}{\langle[gq^2\!+\!a\!+\!b\big({2}/{d_p}\!+\!1\big){P}^2]\delta{P}^2+\!{b}{\delta}P^4/2\!+\!{\lambda}{\delta}P^6/3\!+\!{\lambda}\big({4}/{d_p}\!+\!1\big)({P}^4\delta{P}^2\!+\!{P}^2\delta{P}^4)\!-\!m_p\delta{P}\partial_{\tau}^2\delta{P}\rangle}\Big\}\!+\!{\partial_{P}}S_G\!=\!0\nonumber\\
  &&\Rightarrow\!\!\int\!\!\frac{d{\bf q}}{{(2\pi)^3}}\sum_n\frac{k_BT[{b\big({2}/{d_p}\!+\!1\big){P}\!+\!{\lambda}\big({4}/{d_p}\!+\!1\big)(2{P}^3\!+\!3{P}\langle\delta{P}^2\rangle)}]\langle\delta{P}^2\rangle}{\langle[gq^2\!+\!a\!+\!b\big({2}/{d_p}\!+\!1\big){P}^2\!-\!m_p(i\omega_n)^2]\delta{P}^2+\!{b}{\delta}P^4/2\!+\!{\lambda}{\delta}P^6/3\!+\!{\lambda}\big({4}/{d_p}\!+\!1\big)({P}^4\delta{P}^2\!+\!{P}^2\delta{P}^4)\rangle}\!+\!{\partial_{P}}S_G\!=\!0\nonumber\\
   &&\Rightarrow\!\!\int\!\!\frac{d{\bf q}}{{(2\pi)^3}}\sum_n\frac{k_BT[{b\big({2}/{d_p}\!+\!1\big){P}\!+\!{\lambda}\big({4}/{d_p}\!+\!1\big)(2{P}^3\!+\!3{P}\langle\delta{P}^2\rangle)}]\langle\delta{P}^2\rangle}{[\omega^2_{vm}\big(q,a,b,P^2,\langle\delta{P^2}\rangle\big)-(i\omega_n)^2]\langle\delta{P}^2\rangle}\!+\!{\partial_{P}}S_G\!=\!0\nonumber\\
  &&\Rightarrow\!\![{b\big({2}/{d_p}\!+\!1\big){P}\!+\!{\lambda}\big({4}/{d_p}\!+\!1\big)(2{P}^3\!+\!3{P}\langle\delta{P}^2\rangle)}]\int\frac{d{\bf q}}{(2\pi)^3}\sum_n\frac{k_BT/m_p}{\omega^2_{vm}\big(q,a,b,P^2,\langle\delta{P^2}\rangle\big)-(i\omega_n)^2}\!+\!aP\!+\!bP^3\!+\!\lambda{P^5}\!=\!0\nonumber\\
  &&\Rightarrow\!\![{b\big({2}/{d_p}\!+\!1\big){P}\!+\!{\lambda}\big({4}/{d_p}\!+\!1\big)(2{P}^3\!+\!3{P}\langle\delta{P}^2\rangle)}]\langle\delta{P}^2\rangle\!+\!aP\!+\!bP^3\!+\!\lambda{P^5}\!=\!0\nonumber\\
  &&\Rightarrow\!\!\big[a\!+\!b\big({2}/{d_p}\!+\!1\big)\langle\delta{P}^2\rangle\!+\!3\lambda\big({4}/{d_p}\!+\!1\big)(\langle\delta{P}^2\rangle)^2\big]{P}\!+\!\big[b\!+\!2{\lambda}\big({4}/{d_p}\!+\!1\big)\langle\delta{P}^2\rangle\big]{P}^3\!+\!\lambda{P^5}\!=\!0,
\end{eqnarray}
with
\begin{equation}
\langle\delta{P}^2\rangle=\int\frac{d{\bf q}}{(2\pi)^3}\frac{k_BT}{\hslash}\sum_n\frac{\hslash/m_p}{\omega^2_{vm}\big(q,a,b,P^2,\langle\delta{P^2}\rangle\big)-(i\omega_n)^2}=\int\frac{d{\bf q}}{(2\pi)^3}\frac{\hslash}{m_p}\frac{\coth\big[\hslash\omega_{vm}\big(q,a,b,P^2,\langle\delta{P^2}\rangle\big)/(2k_BT)\big]}{2\omega_{vm}\big(q,a,b,P^2,\langle\delta{P^2}\rangle\big)}. 
\end{equation}
Here, we have taken the variation with respect to $\delta{P}^2$ to obtain thermally averaged $\omega^2_{vm}$ in consideration of the mean-field approximation. Consequently, the thermally averaged fluctuation and the equation to determine the long-range ordered polarization in the main text are derived.

\section{Model for ferroelectric PbTiO$_3$}

We describe the theoretical model of the ferroelectric PbTiO$_3$ in this section. As its crystal structure is tetragonal with a large distortion ratio $1.062$~\cite{lines2001principles,samara1971pressure}, we take the ground-state action of the polarization field limited by the crystal symmetry of PbTiO$_3$~\cite{haun1987thermodynamic}: 
\begin{eqnarray}
  \mathcal{S}\!&=&\int{dtd{\bf x}}\Big\{\frac{m_p}{2}(\partial_t{\bf P})^2-\Big[\frac{g}{2}({\nabla}{\bf P})^2+\frac{a_1}{2}(P_1^2+P_2^2+P_3^2)+\frac{a_{11}}{4}(P_1^4+P_2^4+P_3^4)\!+\!\frac{a_{111}}{6}(P_1^6+P_2^6+P_3^6)+a_{123}P_1^2P_2^2P_3^2\nonumber\\
 &&\mbox{}+a_{112}[P_1^4(P_2^2+P_3^2)+P_2^4(P_1^2+P_3^2)+P_3^4(P_1^2+P_2^2)]+{a_{12}}(P_1^2P_2^2+P_2^2P_3^2+P_1^2P_3^2)\Big]\Big\}.
\end{eqnarray}
Similarly, the polarization field ${\bf P}={\bf P}+\delta{\bf P}$, consisting of the long-range ordered polarization ${\bf P}=P{\bf e}_3$ and polarization fluctuation $\delta{\bf P}=({\delta}P_1{\bf e}_1+{\delta}P_2{\bf e}_2+{\delta}P_3{\bf e}_3)/\sqrt{3}$.    Neglecting the odd orders of the fluctuations, one has    
\begin{align}
 \mathcal{S}=&\!\!\int{dtd{\bf x}}\Big\{\frac{m_p}{2}(\partial_t\delta{P})^2-\Big[\frac{g}{2}({\nabla}\delta{P})^2+\frac{a_1}{2}(P^2+\delta{P}^2)+\frac{a_{11}}{4}\Big(P^4+\frac{{\delta}P^4}{3}+\frac{2P^2\delta{P^2}}{3}\Big)+{a_{12}}\Big(\frac{\delta{P}^4}{3}+\frac{2\delta{P^2}P^2}{3}\Big)\nonumber\\
 &\mbox{}\!+\!\frac{a_{111}}{6}\Big(P^6+\frac{{\delta}P^6}{9}+P^4\delta{P^2}+\frac{P^2\delta{P^4}}{3}\Big)+a_{123}\Big(\frac{\delta{P}^6}{27}+\frac{P^2\delta{P^4}}{9}\Big)+a_{112}\Big(\frac{2\delta{P}^6}{9}+\frac{2P^4\delta{P^2}}{3}+\frac{2P^2\delta{P^4}}{3}\Big)\Big]\Big\},\label{LAGPTO}
\end{align}
where we have considered the classical isotropic fluctuations ($\delta{P}_1=\delta{P}_2=\delta{P}_3=\delta{P}$) and neglect the anisotropic part as an approximation. Consequently, the thermally averaged contributions purely from the long-range ordering in this action gives rise to the free energy in the main text, and the corresponding free-energy parameters are written as
\begin{eqnarray}
  \alpha&=&a_1+\Big(\frac{a_{11}}{3}+\frac{4a_{12}}{3}\Big)\langle\delta{P}^2\rangle+3\Big(\frac{a_{111}}{9}+\frac{4a_{112}}{3}+\frac{2a_{123}}{9}\Big)\langle\delta{P}^2\rangle^2, \\
  \beta&=&a_{11}+\Big(\frac{2a_{111}}{3}+\frac{8a_{112}}{3}\Big)\langle{\delta{P}^2}\rangle, \\
  \lambda&=&a_{111}.  
\end{eqnarray}
Moreover, following the same way, by using the Euler-Lagrange equation of motion with respect to the polarization fluctuation from the action in Eq.~(\ref{LAGPTO}) and employing the thermal field that obeys the fluctuation-dissipation theorem~\cite{landau1981statistical,tang2022excitations}, one can derive the thermally averaged polarization fluctuation:
\begin{equation}
  \langle{\delta{P}^2}\rangle=\int\frac{\hslash}{2m_p}\frac{\coth\big[\hslash\omega^{\rm PTO}_{vm}\big(q,a_{1},a_{11},P^2,\langle\delta{P}^2\rangle\big)/(2k_BT)\big]}{\omega^{\rm PTO}_{vm}\big(q,a_{1},a_{11},P^2,\langle\delta{P}^2\rangle\big)}\frac{d{\bf q}}{(2\pi)^3},  
\end{equation}
where the energy spectrum $\hslash\omega^{\rm PTO}_{vm}\big(q,a_{1},a_{11},P^2,\langle\delta{P}^2\rangle\big)=\hslash\sqrt{\Delta^2_{\rm PTO}\big(a_{1},a_{11},P^2,\langle\delta{P}^2\rangle\big)+gq^2/m_p}$ with the excitation gap:
\begin{eqnarray}
  &&\Delta_{\rm PTO}\big(a_{1},a_{11},P^2,\langle\delta{P}^2\rangle\big)=\frac{1}{\sqrt{m_p}}\Big\{a_1+\Big(\frac{a_{11}}{3}+\frac{4a_{12}}{3}\Big)P^2+\Big(\frac{a_{111}}{3}+\frac{4a_{112}}{3}\Big)P^4\rangle\nonumber\\
  &&\mbox{}+3\Big[\frac{a_{11}}{3}+\frac{4a_{12}}{3}+\Big(\frac{4a_{123}}{9}+\frac{2a_{111}}{9}+\frac{8a_{112}}{3}\Big)P^2\Big]\langle{\delta{P}^2}+15\Big(\frac{a_{111}}{9}+\frac{4a_{112}}{3}+\frac{2a_{123}}{9}\Big)\langle{\delta{P}^2}\rangle^2\Big\}^{1/2}.
\end{eqnarray}
Consequently, following the self-consistent renormalization processes in the main text, one can use the ground-state parameters to produce the dielectric/ferroelectric properties of the ferroelectric PbTiO$_3$ at finite temperatures. For the accuracy, we directly start with the experimentally measured zero-temperature parameters and perform the renormalization by the thermal fluctuations:
\begin{eqnarray}
  \alpha&=&a^{(0)}_1+\Big(\frac{a^{(0)}_{11}}{3}+\frac{4a_{12}}{3}\Big)\langle\delta{P}_{\rm th}^2\rangle+3\Big(\frac{a_{111}}{9}+\frac{4a_{112}}{3}+\frac{2a_{123}}{9}\Big)\langle\delta{P}_{\rm th}^2\rangle^2+\delta\alpha(T), \\
  \beta&=&a^{(0)}_{11}+\Big(\frac{2a_{111}}{3}+\frac{8a_{112}}{3}\Big)\langle{\delta{P}_{\rm th}^2}\rangle, \\
  \lambda&=&a_{111},\\
   \langle{\delta{P}_{\rm th}^2}\rangle&=&\int\frac{\hslash}{2m_p}\Bigg\{\frac{\coth\big[\hslash\omega^{\rm PTO}_{vm}\big(q,a_1^{(0)},a_{11}^{(0)},P^2,\langle{\delta{P}_{\rm th}^2}\rangle\big)/(2k_BT)\big]}{\omega^{\rm PTO}_{vm}\big(q,a_1^{(0)},a_{11}^{(0)},P^2,\langle{\delta{P}_{\rm th}^2}\rangle\big)}-\frac{1}{\omega^{\rm PTO}_{vm}\big(q,a_1^{(0)},a_{11}^{(0)},P_0^2,0\big)}\Bigg\}\frac{d{\bf q}}{(2\pi)^3},  
\end{eqnarray}
where $a^{(0)}_{1}$ and $a^{(0)}_{11}$ are the zero-temperature parameters, and here we have introduced a correction/perturbation term $\delta\alpha(T)$.  The correction due to the coupling of the polarization field with the gapless acoustic phonons through the electrostrictive effect should manifest with characteristic features at low temperatures, but in the ferroelectric phase is masked by the large $1/\varepsilon$ at low~$T$. The correction by this coupling is therefore minimal and can be neglected in the ferroelectrics. However, in the ferroelectric phase, the increased polarization fluctuation with temperature can cause the lattice vibrations and structural instabilities, which influence the harmonic potentials of all lattice degrees of freedom in the self-consistent phonon-field approximation~\cite{verdi2023quantum}. To characterize this effect, we employ the self-consistent field approximation, and the corresponding correction in the ferroelectrics is written as~\cite{maki1986thermal,rice1981impurity}
\begin{equation}
\delta\alpha(T)=-3\eta{a_0}\exp\big[-P^2/\langle{\delta{P_{\rm th}^2(T)}}\rangle\big],  
\end{equation}
with $\eta<1$ being a small dimensionless coefficient. It should be emphasized that $\delta\alpha(T)$ vanishes at $T=0$ and contributes to a small correction only near the phase transition point. 

In the present study, we do not consider the logarithmic corrections~~\cite{roussev2003theory,palova2009quantum} 
 even the system lies in the upper critical dimension. This is because that  in material science, the phase transition of the classical ferroelectric materials, such as PbTiO$_3$ studied in the present work and most classical ferroelectrics (like BaTiO$_3$), is a discontinuous first-order phase transition, where logarithmic corrections are neither needed nor expected. Specifically, logarithmic corrections arise as the subleading modifications to power-law scaling in renormalization group (RG) analysis~\cite{cardy1996scaling}, which are required by scale invariance near a continuous phase transition  governed by a nontrivial fixed point. These corrections become relevant at the upper critical dimension where fluctuations are marginal. However, in a robust first-order transition, there is no divergent correlation length, no critical scaling, and hence no underlying RG fixed point that would mandate such corrections. Our results in the main text further support this theoretical understanding. Without invoking any logarithmic corrections or fitting parameters, our calculation for PbTiO$_3$, using only zero-temperature parameters obtained from independent experiments, yielded an accurate and quantitative description for the experimentally measured both critical behaviors (including transition temperature, first-order-transition nature and mechanism, pyroelectric response) and the full dielectric and ferroelectric properties across the entire temperature range starting from $T=0$.  From a more fundamental viewpoint, conventional renormalization group and scaling theories are primarily designed for continuous (second-order) phase transitions, where divergent correlation lengths and critical fluctuations dominate the physics near the critical point. In contrast, first-order transitions,  characterized by discontinuous changes in the order parameter and finite latent heat as well as non-divergent correlation lengths, generally lie outside the applicability of these frameworks. 

\begin{center}
\begin{table}[htb]
  \caption{Specific model parameters used in our thermodynamic theory for the classical ferroelectric PbTiO$_3$.  The temperature-independent $a_{12}$, $a_{111}$, $a_{112}$ and $a_{123}$ are from Ref.~\cite{haun1987thermodynamic} according to the experimental data at room temperature, whereas the zero-temperature parameters $a_1^{(0)}$ and $a_{11}^{(0)}$ are determined in order to give the experimental estimate $P\sim63~\mu$C/cm$^2$~~\cite{remeika1970growth} and $\varepsilon=167$~\cite{ikegami1971electromechanical} for the zero temperature. $\Delta_{\rm op}$ and $v_{\rm op}$ of the observed optical soft-phonon mode of the ferroelectric state are extracted from the experimental data of the inelastic neutron scattering measurement~\cite{hlinka2006lattice}. The wave-vector cutoff $q_{\rm c,FE}$ in the integral of the bosonic excitation of the collective vector mode is taken as the Debye wave-vector cutoff. The parameter related to the correction term is set as $\eta=0.11$. }\label{PTO}
  \renewcommand\arraystretch{1.5}  
\begin{tabular}{lll}
\hline
\hline
Parameter & Value & Unit \\
\hline
$a_1^{(0)}$ & $-0.1 \times 10^4 / 6.24$ & \AA$\cdot$meV \\

$a_{11}^{(0)}$ & $-3.6 \times 10^7 / (6.24)^3$ & \AA$^5\cdot$meV/e$^4$ \\

$a_{12}$ & $7.5 \times 10^7 / (6.24)^3$ & \AA$^5\cdot$meV/e$^4$ \\

$a_{111}$ & $1.6 \times 10^{12} / (6.24)^5$ & \AA$^9\cdot$meV/e$^6$ \\

$a_{112}$ & $6.1 \times 10^{11} / (6.24)^5$ & \AA$^9\cdot$meV/e$^6$ \\

$a_{123}$ & $-3.66 \times 10^{12} / (6.24)^5$ & \AA$^9\cdot$meV/e$^6$ \\

$v_{\rm op}$ & $52$ & \AA/ps \\

$q_{\rm c,FE}$ & $q_D = (6N\pi^2/\Omega_{\rm cell})^{1/3}$ & -- \\

$\hslash \Delta_{\rm op}$ & $11$ & meV \\
\hline
\hline
\end{tabular}
\end{table}
\end{center}

{\sl Temperature dependence of an ordered ferroelectric phase.---}Here we discuss the temperature dependence of an ordered ferroelectric phase. For a simplified analysis, we use the model in the main text and consider the corresponding zero-temperature parameters $a_0<0$, $b_0\ne0$ and $\lambda\ne0$. Due to the gapped collective vector mode, below the temperature $\hslash\Delta_{\rm op}/k_B\approx127~$K, the thermal fluctuation of the polarization is minimal, leading to the dielectric properties insensitive to the temperature variation. When the increase in temperature exceeds above $\hslash\omega_{vm}(q_c)/k_B$, with $q_c$ being the wave-vector cutoff, the excitation gap $\Delta$ persists, and one finds approximately
\begin{eqnarray}
\langle{\delta{P}_{\rm th}^2}\rangle\!\approx\!k_BT\int_0^{q_c}\frac{2{\hslash}q^2dq/{(2\pi)^2}}{m_p\Delta^2+gq^2}=k_BT\Big[\sqrt{gq_c^2}-\sqrt{m_p\Delta^2}\arctan\Big(\sqrt{gq_c^2/(m_p\Delta^2)}\Big)\Big]\frac{\hslash}{g^{3/2}{2\pi^2}}.~~
\end{eqnarray}
For ${gq_c^2/(m_p\Delta^2)}\gg1$ as usually observed in the ferroelectric materials, $\langle{\delta{P}_{\rm th}^2}\rangle$ exhibits the linear-$T$ behavior, leading to the widely reported classical linear-$T$ dependencies in $1/\varepsilon(T)$ and $\alpha(T)$ in the literature.

\section{Model for quantum paraelectric SrTiO$_3$ and KTaO$_3$}

The structures of SrTiO$_3$ and KTaO$_3$ are found to be cubic perovskite  at room temperature.  While KTaO$_3$ remains as cubic down to low-temperature limit~\cite{singh1996stability},  SrTiO$_3$ undergoes a cubic-to-tetragonal antiferrodistortive transition at $105~$K~\cite{shirane1969lattice,vogt1995refined,cowley1996phase}, but its tetragonality of $1.00056$ is very small, causing essentially no change in the cell volume, thermal-expansion coefficient, or dielectric properties at the antiferrodistortive transition~\cite{beattie1971pressure,tao2016nonmonotonic}.
 Thus, we can directly use the model developed in Sec.~\ref{III}.

\subsection{Correction of acoustic phonon in quantum paraelectrics SrTiO$_3$ and KTaO$_3$}

In the quantum paraelectrics with a nearly vanishing $a_0$, the correction due to the coupling of the polarization field with the gapless acoustic phonons through the electrostrictive effect, which should manifest with the characteristic features at low temperatures, becomes inevitable because of the small $1/\varepsilon$ at low-temperature limit.  In this part, we introduce a model to include this coupling~\cite{palova2009quantum,khmelnitskii1973phase,rowley2014ferroelectric} and its correction to the renormalization processes. Specifically, the electrostrictive effect originates from the three-phonon interactions between two optical phonons and one acoustic phonon, and results in a coupling between the long-range ordered polarization and strain $s_i$ (voigt notation)~\cite{haun1987thermodynamic}. This coupling energy in cubic perovskites reads $E_{\rm int}=-C\sum_{i=1,2,3}s_i{P^2}$ with $C>0$ being the cubic electrostrictive constant. By adding this coupling contribution to the free energy of the long-range ordered polarization, one directly finds a thermally averaged correction:
\begin{equation}\label{dadada}
\delta\alpha=-C\langle{s}\rangle,  
\end{equation}
where $\langle{s}\rangle=\langle{s_1}\rangle=\langle{s_2}\rangle=\langle{s_3}\rangle$ (cubic perovskite), and this thermal average within the quantum statistic theory of the acoustic phonon is given by
\begin{equation}\label{SS}
\langle{s}\rangle=-\sum_{\lambda}\frac{d_{\lambda}}{{\rho}v_{\lambda}^2}\int\frac{\hslash{q^2}}{\rho}\frac{n_B(\hslash{v_{\lambda}q})}{2v_{\lambda}q}\frac{d{\bf q}}{(2\pi)^3}.  
\end{equation}
Here, $\rho$ stands for the mass density; $v_{\lambda}$ denotes the group velocity of the $\lambda$-branch (LA,TA) acoustic phonon;  $d_{\lambda}<0$ represents the contribution coefficient from the $\lambda$-branch acoustic phonon to the thermal expansion, which is related to the third-order non-harmonic coefficient of the strain. Consequently, by adding the contributions from the above equations in the self-consistent formulation of the renormalization processes by the thermal fluctuations, the coupling effect of the polarization field with acoustic phonons via electrostrictive effect is incorporated in the our thermodynamic theory. 

{\sl Detailed derivations.---}We next present the detailed derivations of Eq.~(\ref{SS}). We start from the action of the strain (i.e., acoustic phonons):
\begin{equation}
\mathcal{S}_{s}=\int{dt{d{\bf x}}}\Big[\frac{1}{2}\rho(\partial_t{\bf u}_{\rm LA})^2\!-\!\frac{1}{2}e_{\rm LA}s^2_{\rm LA}\!-\!\frac{1}{3}d_{111}{s_{\rm LA}^3}\!-\!3d_{144}s_{\rm LA}s^2_{\rm TA}\!+\!\frac{1}{2}\rho(\partial_t{\bf u}_{\rm TA})^2\!-\!\frac{1}{2}e_{\rm TA}s^2_{\rm TA}-\frac{1}{3}d_{444}{s_{\rm TA}^3}\!-\!3d_{114}s^2_{\rm LA}s_{\rm TA}\Big],
\end{equation}
where ${\bf u}_{\lambda}$ and $s_{\lambda}$ denote the lattice vibration and corresponding strain contributed by the $\lambda$-branch acoustic phonon, respectively; $e_{\lambda}$ and $d_{ijk}$ are the second-order harmonic and third-order non-harmonic coefficients. Using the Euler-Lagrange
equation of motion~\cite{peskin2018introduction}, one finds
\begin{eqnarray}\label{sla}
  D^{-1}_{\rm LA}(t,{\bf q})s_{\rm LA}&=&\sigma_{\rm LA}(t,{\bf q})-d_{\rm 111}s_{\rm LA}^2-3d_{144}s_{\rm TA}^2-6d_{114}s_{\rm LA}s_{\rm TA},\\
  D^{-1}_{\rm TA}(t,{\bf q})s_{\rm TA}&=&\sigma_{\rm TA}(t,{\bf q})-d_{\rm 444}s_{\rm TA}^2-3d_{114}s_{\rm LA}^2-6d_{144}s_{\rm TA}s_{\rm LA},\label{sta}
\end{eqnarray}
where the inverse Green function $D^{-1}_{\lambda}(t,{\bf q})=(\rho/q^2)\partial_t^2+e_{\lambda}+\gamma\partial_t$ with $\gamma=0^+$ being the damping rate~\cite{abrikosov2012methods}. Here, we have introduced the thermal field $\sigma_{\lambda}(t,{\bf q})$ which obeys the fluctuation-dissipation theorem~\cite{landau1981statistical,tang2022excitations}:
\begin{eqnarray}\label{fdt}
\langle\sigma_{\lambda}(\omega,{\bf q})\sigma^*_{\lambda'}(\omega',{\bf q}')\rangle=\frac{\gamma{\hslash\omega}(2\pi)^4\delta(\omega-\omega')\delta({\bf q}-{\bf q}')}{\tanh[\hslash\omega/(2k_BT)]}\delta_{\lambda,\lambda'}.
\end{eqnarray}  
From Eqs.~(\ref{sla}) and~(\ref{sta}), keeping the leading terms, one has
\begin{equation}
s_{\rm LA}=D_{\rm LA}\sigma_{\rm LA}-d_{\rm 111}D_{\rm LA}(D_{\rm LA}\sigma_{\rm LA})^2-3d_{144}D_{\rm LA}(D_{\rm TA}\sigma_{\rm TA})^2-6d_{114}D_{\rm LA}(D_{\rm LA}\sigma_{\rm LA})(D_{\rm TA}\sigma_{\rm TA}).
\end{equation}
Consequently, with Eq.~(\ref{fdt}), the thermal average is derived as
\begin{equation}
\langle{s_{\rm LA}}\rangle=-\frac{d_{111}}{e_{\rm LA}}\int\frac{\hslash{q^2}}{\rho}\frac{n_B(\hslash{q\sqrt{e_{\rm LA}/\rho}})}{2q\sqrt{e_{\rm LA}/\rho}}\frac{d{\bf q}}{(2\pi)^3}-\frac{3d_{144}}{e_{\rm TA}}\int\frac{\hslash{q^2}}{\rho}\frac{n_B(\hslash{q\sqrt{e_{\rm TA}/\rho}})}{2q\sqrt{e_{\rm TA}/\rho}}\frac{d{\bf q}}{(2\pi)^3}.    
\end{equation}
Here, we have neglected the zero-point oscillations of the acoustic phonons. Considering the group velocity of the $\lambda$-branch acoustic phonon $v_{\lambda}=\sqrt{e_{\lambda}/\rho}$ and $\langle{s_{\rm LA}}\rangle=\langle{s_1}\rangle=\langle{s}\rangle$ for the cubic perovskite, one arrives at Eq.~(\ref{SS}).  \\

Recent theoretical~\cite{PhysRevLett.131.046801,PhysRevB.89.184104} and experimental~\cite{zhang2025nanoscale,PhysRevB.106.L140301} studies have suggested that flexoelectric couplings, arising from spatial gradients of the polarization field coupled to strain, can soften the transverse acoustic mode  and enhance spatially modulated fluctuations of the polar field in quantum paraelectrics. While such effects are expected to influence domain-wall structures, nanoscale inhomogeneity, or systems with strong strain gradients, it is important to note that flexoelectric interactions primarily modify higher-momentum ($q\neq0$) components of the polarization (order parameter), and since the flexoelectric coupling is linear in the polarization gradient, it does not renormalize the collective-mode energy spectrum or its excitation gap.  In contrast, the macroscopic dielectric function probed in experiments is governed by the long-wavelength ($q\rightarrow 0$) component of the inverse susceptibility. Our work focuses precisely on this global, long-range-ordered ($q\sim0$) polarization behavior, which determines the experimentally measured bulk dielectric properties. Therefore, although flexoelectric couplings can in principle generate spatially modulated polarization fluctuations or influence nanoscale structure, their contribution to the global polarization and to the experimentally measured macroscopic $q=0$ response is expected to be negligible.
  Conversely, the present computational framework provides a natural foundation for systematically incorporating flexoelectric couplings in future study, without relying on phenomenological Ginzburg-Landau constructions as done previously. Once this basis is accurately established, gradient-induced flexoelectric terms can be introduced to formulate and describe nanoscale patterns, inhomogeneous strain environments, and spatially modulated polarization states.

\subsection{Temperature dependence of inverse dielectric function in quantum paraelectrics}

In this part, we present an analytical analysis of the temperature dependence of the inverse dielectric function in the quantum paraelectrics. In the quantum paraelectrics SrTiO$_3$ and KTaO$_3$, as the inverse dielectric function $1/\varepsilon(T)\propto\alpha(T)$, one finds
\begin{equation}
1/\varepsilon(T)\propto{a_0+b_0(2/d_p+1)\langle{\delta{P}_{\rm th}^2}\rangle}+\delta\alpha.   
\end{equation}

At low temperatures near zero, due to the gapped collective vector mode and hence minimum $\langle{\delta{P}_{\rm th}^2}\rangle$, the temperature variation of $1/\varepsilon(T)$ is dominated by the correction term $\delta\alpha$, i.e, the thermal expansion associated with the gapless acoustic phonon via the electrostrictive effect. It therefore decreases with the increase of temperature according to Eq.~(\ref{dadada}). 

With further increase in temperature, the increased thermal fluctuation from zero starts to dominate the temperature variation of $1/\varepsilon(T)$. In this situation, considering $P\equiv0$ and nearly vanishing $a_0$ in quantum paraelectrics but neglecting the minimal $\langle{\delta{P}_{\rm th}^2}\rangle$ in the excitation gap $\Delta$, one approximately has
\begin{eqnarray}
\langle{\delta{P}_{\rm th}^2}\rangle\!\approx\!\!\int\frac{{\hslash}n_B\big(\hslash{q}\sqrt{g/m_p}\big)q^2dq}{2\pi^2m_pq\sqrt{g/m_p}}\!=\!\!\int\frac{\hslash{n_B(\omega)}{\omega}d\omega}{2\pi^2m_p({g/m_p})^{3/2}}\!=\!-\!\sum_{n=1}^{\infty}\frac{\hslash(k_BT)^2(1\!+\!nx)e^{-nx}\big|_{x=0}^{x=\frac{\hslash\omega_c}{k_BT}\approx\infty}}{2\pi^2n^2m_p({g/m_p})^{3/2}}\!=\!\frac{\hslash(k_BT)^2\zeta(2)}{2\pi^2m_p({g/m_p})^{3/2}},~~~~
\end{eqnarray}
leading to a non-classical $T$-square temperature behavior of the inverse dielectric function. 

When the temperature increases to exceed $\hslash\omega_{vm}(q_c)/k_B$ at high temperatures, with $q_c$ being the wave-vector cutoff, one approximately has
\begin{eqnarray}
\langle{\delta{P}_{\rm th}^2}\rangle\!\approx\!k_BT\int_0^{q_c}\frac{2{\hslash}q^2dq/(2\pi)^2}{b_0\langle{\delta{P_{\rm th}^2}}\rangle+gq^2}=k_BT\Big[\sqrt{gq^2_c}-\sqrt{b_0\langle{\delta{P_{\rm th}^2}}\rangle}\arctan\Big(\sqrt{gq_c^2/b_0\langle{\delta{P_{\rm th}^2}}\rangle}\Big)\Big]\frac{{\hslash}}{2\pi^2g^{3/2}}.~~\label{thhighT}
\end{eqnarray}
For SrTiO$_3$ and KTaO$_3$, $gq^2_c/b_0\langle{\delta{P_{\rm th}^2}}\rangle\approx{4g\varepsilon(T)q^2_c}\gg1$, thereby leading to $\langle{\delta{P}_{\rm th}^2}\rangle\propto{T}$, and hence, the inverse dielectric function exhibits the classical linear-$T$ (i.e., Curie–Weiss) behavior at high temperatures.

\begin{center}
\begin{table}[htb]
  \caption{Specific parameters used in our thermodynamic theory for the quantum paraelectric SrTiO$_3$.  For the model parameters at zero temperature, $v_{\rm op}$ and $\Delta_{\rm op}$ of the optical soft-phonon mode of the incipient ferroelectric state are from Ref.~\cite{rowley2014ferroelectric} by comparing the data from inelastic neutron~\cite{yamada1969neutron} and Raman scattering~\cite{fleury1968electric} experiments at 4~K, and $a_0$ and $b_0$ are from Ref.~\cite{rowley2014ferroelectric} by measuring  $E/P=a_0+b_0P^2$ at 0.3~K, with $E$ being the electric field (0 to 15~kV/cm) and $P$ denoting the electric polarization. The wave-vector cutoff $q_{\rm c,FE}$ in the integral for the bosonic excitation of the collective vector mode, which determines the Curie–Weiss behavior as shown by Eq.~(\ref{thhighT}), is extracted from the Curie constant in the experimental data~\cite{rowley2014ferroelectric}. As for the correction from the acoustic phonons,  the group velocity of the acoustic phonon modes, $v_{\rm LA}$ (longitudinal acoustic mode) and $v_{\rm TA,(1)}$ (transverse acoustic mode) are given in Ref.~\cite{kor1975phonon} by experimental measurement. Moreover, Ref.~\cite{he2020anharmonic} reported the disappearance of a transverse acoustic phonon branch in SrTiO$_3$ at low temperatures, due to the coupling with the optical phonon of the incipient ferroelectric instability near the quantum critical point, and this mode is sometimes referred to as the second sound in the literature.  Here we take account of this anomalous disappearance by, as an approximation, considering a softening of this acoustic phonon branch, i.e., an effectively reduced group velocity $v_{\rm TA, (2)}$,  which is determined by taking the recently reported/estimated group velocity from the thermal-conductivity measurement near the tetragonal phase of SrTiO$_3$~\cite{zhang2022thermal}.  The cubic electrostrictive constant $C$ is determined by the first-principles method and it shows good agreement with the experimental data of strain~\cite{schimizu1997effect}. The wave-vector cutoff $q_{\rm c,AC}$ in the integral of the acoustic phonons is chosen to fit the minimum (single point) of the dielectric function at the low-temperature limit. We noticed that there could be a hot effect during the measurement in Ref.~\cite{rowley2014ferroelectric}, and in consideration of this effect, the experimental data of SrTiO$_3$ in Fig.~2 in the main text have been shifted by 1.66~K toward the high-temperature direction. }  
\renewcommand\arraystretch{1.5}   \label{STO}
\begin{tabular}{l l l l}
    \hline
    \hline
   model parameters &~~~&\quad\quad\quad\quad\quad~~acoustic-phonon correction &~~~~\\
    \hline
    $a_0~$(meV$\cdot${\AA}/e$^2$)&\quad~$5.061\times10^{-5}/\varepsilon_0$&\quad\quad\quad\quad\quad~~$v_{\rm LA}~$({\AA}/ps)&\quad~78.66 \\
    $b_0~$(meV$\cdot${\AA}$^5$/e$^4$)&\quad~$0.07\times(2\pi)^3/\varepsilon_0$&\quad\quad\quad\quad\quad~~$v_{\rm TA,(1)}~$({\AA}/ps)&\quad~49.04\\
    $\hslash\Delta_{\rm op}/k_B$~(K)&\quad~$24$&\quad\quad\quad\quad\quad~~$v_{\rm TA,(2)}~$({\AA}/ps)&\quad~10\\
    $v_{\rm op}~$({\AA}/ps)&\quad~$81$&\quad\quad\quad\quad\quad~~$d_{111}~$(meV/{\AA}$^3$)&\quad~$-5\times10^4/1.6$\\
    $q_{\rm c,FE}$~({\AA}$^{-1}$)&\quad~1.15&\quad\quad\quad\quad\quad~~$3d_{144}~$(meV/{\AA}$^3$)&\quad~$-2\times10^4/1.6$\\
    &\quad~&\quad\quad\quad\quad\quad~~$C~$(meV$\cdot${\AA}/e$^2$)&\quad~24.08$a_0$\\
    &\quad~&\quad\quad\quad\quad\quad~~$q_{\rm c,AC}$~({\AA}$^{-1}$)&\quad~0.15\\
       \hline
    \hline
\end{tabular}\\
\end{table}
\end{center}

\subsection{Renormalization by zero-point fluctuations}

We show in this part that the renormalization by the zero-point fluctuations on the bare parameters is important for accurate predictions as compared with the experimental measurements at low-temperature limit. In principle, the formulation of the bare parameters in the displacive ferroelectrics is usually associated with the structural/lattice instabilities. It therefore requires the first-principles calculation with highly accurate computation of the self-consistent harmonic lattice dynamics and higher-level treatment of the electronic correlation. Since similar formulation has recently been performed in SrTiO$_3$~\cite{verdi2023quantum}, we take SrTiO$_3$ as the example here. The first-principles calculation adopting the PBEsol/rSCAN functional in Ref.~\cite{verdi2023quantum} reported the unstable phonon modes with imaginary frequencies (i.e., negative values of the harmonic terms in lattice dynamics) at the $\Gamma$ point in cubic phase of SrTiO$_3$. According to the lattice dynamics and condensation of the imaginary-frequency phonons  ($\omega_{\rm im}^2=-\Delta_{\rm im}^2+v^2_{\rm op}q^2$), one has $a=-{\nu}m_p\Delta^2_{\rm im}$, with $\nu<1$ being the condensation ratio, and hence, the imaginary frequencies result in a negative value of the bare parameter $a$. For SrTiO$_3$, with the calculated $\hslash\Delta_{\rm im}\approx7.7~$meV from PBEsol functional~\cite{verdi2023quantum}, one has $a=-70.15\times10^{-5}/\varepsilon_0$, suggesting a ferroelectric ground state of SrTiO$_3$ at the bare case. Then, from the self-consistent renormalization by the zero-point fluctuations in the developed thermodynamic theory, taking the wave-vector cutoff in the integral of the  zero-point oscillations of the collective vector mode as the Debye wave-vector cutoff $q_{D}=(6N\pi^2/\Omega_{\rm cell})^{1/3}=1.7~$\AA$^{-1}$ for the quantum fluctuations, our numerical calculation predicts $a_{0,{\rm zo}}\approx3.21\times10^{-5}/\varepsilon_0$, in good agreement with the experimentally measured one $a_0\approx5.06\times10^{-5}/\varepsilon_0$ (Table~\ref{STO}). In other words, because of the existence of the unstable phonon mode, SrTiO$_3$ should undergo a transition to the ferroelectric phase at low temperatures, but the zero-point oscillation/vibration of the collective vector mode prevents the formation of the long-range ferroelectric ordering, leading to an incipient ferroelectricity (quantum paraelectric state). This theoretical/analytical description confirms the established understanding of the quantum paraelectric ground state of SrTiO$_3$ in the literature, while Ref.~\cite{verdi2023quantum} has also reported that when the self-consistent quantum anharmonic fluctuations are accounted for in the first-principles calculation, the original unstable phonon modes at the $\Gamma$ point are found to be stable. Consequently, in the quantum paraelectrics, the bare state is ferroelectric but the zero-temperature state is paraelectric. Because of this unique character, the collective vector mode can persist to exist and be uninterrupted in a wide temperature range above zero temperature. This also leads to an interesting open question in the further investigation that whether the ground state of the quantum paraelectrics can be interpreted as a state that contains the creation and annihilation of the many-particle condensation, similar to the quantum-field vacuum state that contains the creation and annihilation of the various single particles.

\begin{center}
\begin{table}[htb]
  \caption{Specific parameters used in our thermodynamic theory for the quantum paraelectric KTaO$_3$.  For the model parameters at zero temperature, $v_{\rm op}=57~${\AA}/ps and $\hslash\Delta_{\rm op}/k_B=36~$K of the optical soft-phonon mode of the incipient ferroelectric state were determined in Ref.~\cite{rowley2014ferroelectric} by comparing the data from inelastic neutron~\cite{shirane1967temperature} and Raman scattering~\cite{fleury1968electric} experiments at 4~K.  In the simulation for each experiment, $a_0$ is extracted from the measured $1/\varepsilon(T)$ near $T=0$, and $b_0$ is determined by $1/\varepsilon(T=10~K)$.  The wave-vector cutoff $q_{\rm c,FE}$ in the integral for the bosonic excitation of the collective vector mode, which determines the Curie–Weiss behavior as shown by Eq.~(\ref{thhighT}), is extracted from the Curie constant in each experiment. As for the correction from the acoustic phonons, the group velocity of the acoustic phonon modes, $v_{\rm LA}$ and $v_{\rm TA,(1)}$, are given in Ref.~\cite{xu2017first} through various approaches, and the second-sound velocity $v_{\rm TA,(2)}$ comes from Refs.~\cite{koreeda2010collective,farhi2000extra,farhi2000broad}. The cubic electrostrictive constant $C$ is determined by the experimental finding that the strain-induced ferroelectric phase transition occurs at a critical stress $\sigma_c=5.5\times10^{9}$dyn/cm$^2$~\cite{uwe1975electrostriction}, i.e., $C=(1/s_c)a_0$ with $s_c$ being the critical strain. The strain coefficients $c_{11}$, $d_{111}$ and $d_{144}$ were addressed in Ref.~\cite{achar1981static}.  The wave-vector cutoff $q_{\rm c,AC}$ in the integral of the acoustic phonons is chosen to fit the minimum (single point) of the dielectric function at low-temperature limit in Ref.~\cite{rowley2014ferroelectric}. We noticed that there could be a hot effect during the measurement in Ref.~\cite{rowley2014ferroelectric}, and in consideration of this effect, the experimental data of KTaO$_3$ in Fig.~3 in the main text have been shifted by 2.5~K toward the high-$T$ direction. }  
\renewcommand\arraystretch{1.5}  \label{KTO}
\begin{tabular}{l l l l l l}
    \hline
    \hline  
    model parameters&\quad$a_0~$(meV$\cdot${\AA}/e$^2$)&\quad$b_0~$(meV$\cdot${\AA}$^5$/e$^4$)&\quad\quad$q_{\rm c,FE}$~({\AA}$^{-1}$)\\
    for EXP in Ref.~\cite{rowley2014ferroelectric}&\quad $22\times10^{-5}/\varepsilon_0$   &\quad $0.24\times(2\pi)^3/\varepsilon_0$  &\quad\quad\quad 0.7\\
    for EXP in Ref.~\cite{wemple1965some}&\quad $22.7\times10^{-5}/\varepsilon_0$ &\quad $0.3\times(2\pi)^3/\varepsilon_0$  &\quad\quad\quad 0.6\\
    for EXP in Ref.~\cite{abel1971effect}&\quad $25\times10^{-5}/\varepsilon_0$   &\quad $0.3\times(2\pi)^3/\varepsilon_0$  &\quad\quad\quad 0.57\\
    for EXP in Ref.~\cite{ang2001dielectric}&\quad $23\times10^{-5}/\varepsilon_0$   &\quad$0.3\times(2\pi)^3/\varepsilon_0$ &\quad\quad\quad 0.67\\
    for EXP in Ref.~\cite{aktas2014polar}&\quad $20.1\times10^{-5}/\varepsilon_0$ &\quad $0.16\times(2\pi)^3/\varepsilon_0$ &\quad\quad\quad 0.62\\  
    \hline
    acoustic-phonon correction&\quad~$v_{\rm LA}$&\quad\quad\quad~$v_{\rm TA,(1)}$&\quad\quad\quad~$v_{\rm TA,(2)}$&\quad\quad~$C$~(meV$\cdot${\AA}/e$^2$) \\
   &\quad~75&\quad\quad\quad~44&\quad\quad\quad~10&\quad\quad\quad~${22a_0}$\\
        &$d_{111}~$(meV/{\AA}$^3$)&\quad~$3d_{144}~$(meV/{\AA}$^3$)&\quad\quad~$q_{\rm c,AC}$~({\AA}$^{-1}$)&\quad\quad~$c_{11}~$(meV/{\AA}$^3$)\\
     &$-5.5\times10^4/1.6$&\quad~$-1.2\times10^4/1.6$&\quad\quad\quad0.22&\quad\quad\quad0.37$\times10^4$/1.6\\       
    \hline
       \hline
\end{tabular}\\
\end{table}
\end{center}

\section{Parameters used in the numerical simulation based on our thermodynamic theory for the renormalization by thermal fluctuations}

In this part, we present the specific values of the used zero-temperature parameters in the simulation and the way in which they were determined. According to the lattice dynamics of the unstable phonon mode, the polarization inertia is given by $m_p=\frac{\Omega_{\rm cell}}{\sum_iQ_i^2/M_i}$~\cite{sivasubramanian2004physical,tang2022excitations},  where $M_i$ and $Q_i$ denote the ionic masses and charges in the unit cell of volume $\Omega_{\rm cell}$, respectively. Moreover, because of the lattice dynamics, the coefficient $g$ is related to the velocity $v_{\rm op}$ of the imaginary-frequency phonons ($\omega_{\rm im}^2=-\Delta_{\rm im}^2+v^2_{\rm op}q^2$) of the ferroelectrics and is written as ${g}/{m_p}=v^2_{\rm op}\times\Delta^2(a_0,b_0,P_0^2,0)/{\Delta^2_{\rm op}}$ following the treatment in Ref.~\cite{rowley2014ferroelectric} as a consequence of the condensation. Here, $\Delta_{\rm op}$ denotes the excitation gap of the experimentally observed optical soft phonon mode ($\omega_{\rm op}^2=\Delta_{\rm op}^2+v^2_{\rm op}q^2$) of the ferroelectric state at low-temperature limit. Other used specific parameters in the developed thermodynamic theory for the classical ferroelectric PbTiO$_3$ as well as quantum paraelectric SrTiO$_3$ and KTaO$_3$ are addressed in Tables~\ref{PTO} as well as~\ref{STO} and \ref{KTO}, respectively. 

{It should be emphasized that the key parameters ($a_0$, $b_0$, $\Delta_{\rm op}$, and $v_{\rm op}$ as well as $q_{c,{\rm FE}}$) governing the main ferroelectric and paraelectric properties in our model are consistently determined from low-temperature experimental inputs/behaviors. Only for the smaller correction terms discussed above (e.g., those contributing to the upturn in the insets of Fig.~3), certain parameters, such as those associated with acoustic phonons and their coupling to the polarization field,  are not directly available from low-temperature experiments. In these cases, we rely on a combination of first-principles calculations and available experimental data. These corrections have only a minor quantitative effect on the main results. Therefore, the predictive capability of the model, particularly for the dominant dielectric and ferroelectric behavior, is not affected by this mixed sourcing of parameters.}

\end{appendix}

\end{widetext}

%

\end{document}